\documentclass[final]{siamltex}

\usepackage{epsfig,amssymb,latexsym}
\usepackage{amsfonts,psfrag,amsmath,bbm,color}
\usepackage{cancel}
\usepackage{comment}
\usepackage{mathrsfs}
\usepackage{graphicx}
\usepackage{textcomp}
\usepackage{multirow}
\usepackage{enumerate}
\usepackage{cancel}
\usepackage{algpseudocode}
\usepackage{caption}  
\usepackage{subcaption}
\usepackage{url}
\usepackage{rotating}
\usepackage{slashbox}
\usepackage{bigints}
\usepackage{hyperref}
\usepackage{xcolor}
\usepackage{placeins}

\newcommand{\bnh}{\hat{\textbf{\textit{n}}}}

\newcommand{\bJ}{{\bf{J}}}
\newcommand{\bE}{{\bf E}}
\newcommand{\bs}{{\bf s}}

\vbadness=\maxdimen

\newcommand{\by}{\boldsymbol{y}}
\newcommand{\footremember}[2]{%
    \footnote{#2}
    \newcounter{#1}
    \setcounter{#1}{\value{footnote}}%
}
\newcommand{\footrecall}[1]{%
    \footnotemark[\value{#1}]%
} 

\usepackage{tikz}
\usetikzlibrary{shapes.geometric, arrows}
\tikzstyle{startstop} = [rectangle, rounded corners, minimum width=1cm, minimum height=1cm,text centered, draw=black]
\tikzstyle{io} = [trapezium, trapezium left angle=70, trapezium right angle=110, minimum width=1cm, minimum height=1cm, text centered, draw=black, fill=blue!30]
\tikzstyle{method} = [rectangle, rounded corners, minimum width=1cm, minimum height =1cm, text centered, draw=black]
\tikzstyle{process} = [rectangle, minimum width=1cm, minimum height=1cm, text centered, draw=black]
\tikzstyle{decision} = [diamond, minimum width=0.5cm, minimum height=0.5cm, text centered, draw=black, fill=green!30]
\tikzstyle{arrow} = [thick,->,>=stealth]

\usepackage{amscd}

\graphicspath{{./figs/}}

\def\PP{{{\rm l}\kern - .15em {\rm P} }}
\def\PN2{{\PP_{N}-\PP_{N-2}}}

\newcommand{\bb}{\boldsymbol{b}}
\newcommand{\bB}{\boldsymbol{B}}

\newcommand{\bH}{\boldsymbol{H}}

\newcommand{\bx}{\boldsymbol{x}}

\newcommand{\deleted}[1]{{}}

\begin{document}
\title{SCALABILITY ANALYSIS OF DIRECT AND ITERATIVE SOLVERS USED TO MODEL CHARGING OF NON-INSULATED SUPERCONDUCTING PANCAKE SOLENOIDS}

\author{
 M. Mohebujjaman\footnote{C\MakeLowercase{orrespondence: jaman@psfc.mit.edu}}\footremember{mit}{P\MakeLowercase{lasma} S\MakeLowercase{cience and} F\MakeLowercase{usion} C\MakeLowercase{enter}, M\MakeLowercase{assachusetts} I\MakeLowercase{nstitute of} T\MakeLowercase{echnology}, C\MakeLowercase{ambridge}, MA 02139, USA}%
\and S. Shiraiwa\footremember{pppl}{P\MakeLowercase{rinceton} P\MakeLowercase{lasma} P\MakeLowercase{hysics} L\MakeLowercase{aboratory}, P\MakeLowercase{rinceton} U\MakeLowercase{niversity}, P\MakeLowercase{lainsboro} T\MakeLowercase{ownship}, NJ 08536, USA}%
\and B. LaBombard\footrecall{mit}%
\and J. C. Wright\footrecall{mit}
 \and K. Uppalapati\footremember{csf}{C\MakeLowercase{ommonwealth} F\MakeLowercase{usion} S\MakeLowercase{ystems}, C\MakeLowercase{ambridge}, MA 02139, USA}%
 }
\maketitle

\begin{abstract}
A mathematical model for the charging simulation of non-insulated superconducting pancake solenoids \textcolor{black}{is presented.} \textcolor{black}{Numerical solutions are obtained by the simulation model implemented on the Petra-M FEM platform using a variety of solvers.}  A scalability analysis is performed for \textcolor{black}{both} direct and preconditioned iterative solvers \textcolor{black}{for} four different pancakes solenoids with varying number of turns and mesh elements. It is found that \textcolor{black}{even with two extremely different time scales in the system} an iterative solver combination (FGMRES-GMRES) in conjunction with the parallel Auxiliary Space Maxwell Solver \textcolor{black}{(AMS)}  preconditioner outperforms \textcolor{black}{a parallelized} direct solver (MUMPS). In general, the computational time \textcolor{black}{of the iterative solver} \textcolor{black}{is found to} increase \textcolor{black}{with the} number of turns \textcolor{black}{in the solenoids} and/or the conductivity \textcolor{black}{assumed for the superconducting material.}  
\end{abstract}


\medskip

\pagestyle{myheadings}
\thispagestyle{plain}
\markboth{M. Mohebujjaman, S. Shiraiwa, B. LaBombard, J. C. Wright, and K. Uppalapati}{SCALABILITY ANALYSIS OF DIRECT AND ITERATIVE SOLVERS}

\section{Introduction}
The discovery of high-temperature superconductors (HTS) \cite{ginsberg1992physical, liu2017analysis} opens a new chapter in both scientific and engineering fields for producing high-field superconducting magnets. \textcolor{black}{Non-insulated superconductors in particular show great promise \cite{hahn2010hts}. However, } the numerical simulation of \textcolor{black}{a non-insulated} superconducting magnet is computationally expensive due to its highly non-linear electromagnetic behavior \cite{ruiz2004computer}. Though for simple geometries, the analytical solutions are possible to find under uniform external magnetic field \cite{Brandt1996Superconductors, Prigozhin1996Bean}, time-dependent magnetic field simulation still remains a challenge to the scientific community even with the advanced computing facilities. This is because of (i) huge simulation domain as it includes air region  \textcolor{black}{along with the magnet that can produce multi-billions degrees of freedom (dofs) even taking advantage of symmetries (ii) complex magnet and conductor path geometry (iii) materials with widely disparate conductivity that can include an air region, $\sigma_{air}=1.0\hspace{1mm} S/m$, in direct contact with a superconducting region, $\sigma_{hts}=10^{15}\hspace{1mm}S/m$, producing a system matrix in a finite element discretization of the PDEs that can be highly ill-conditioned,} and (iv) null-space of the $curl$-$curl$ operator in the governing Maxwell equations. \textcolor{black}{Consider a sparse linear system
\begin{align}
 A\bx=\bb\hspace{2mm}\text{with}\hspace{2mm} A\in\mathbb{R}^{N\times N},\label{system1}
\end{align}
 where $N$ is the order of the matrix $A$}. The sparse direct solvers usually compute the $LU$ decomposition (or its variant) of the system matrix $A$ using its sparsity pattern in an efficient way so that it becomes easier to compute and store the factors. The direct solver \textcolor{black}{inverts} the system exactly, provides a very robust solution, \textcolor{black}{and easy to use}.  For solving a typical PDE using a \textcolor{black}{sparse} direct solver, the computational cost grows as $O(N^2)$ in 2D and $O(N^{7/3})$ in 3D, and memory requirement grows as $O(N\log N)$ in 2D and $O(N^{4/3})$ in 3D \cite{bangerthLecture34, langer2018direct}. Over the last few decades, due to the massive improvement \textcolor{black}{in} the sparse direct solvers, e.g., MUltifrontal Massively Parallel sparse direct Solver (MUMPS)\footnote{http://mumps.enseeiht.fr/} \cite{amestoy2000multifrontal, amestoy2000mumps}, \textcolor{black}{PARDISO\footnote{https://www.pardiso-project.org/} \cite{kuzmin2013fast}, STRUMPACK \cite{ghysels2016efficient,rouet2016distributed},}  SuperLU \cite{li2011superlu}, or UMFPACK  \cite{davis2004algorithm} \textcolor{black}{have} become popular among the researchers. Notwithstanding the advent of robust direct solvers, as the problem size and complexity increase, they require \textcolor{black}{increasing computational time and computer memory and thus often fail to produce sufficiently resolved long range time-dependent solutions due to the limit of simulation time and memory constraint.}
Iterative solvers, e.g., CG \cite{hestenes1952methods}, BiCGSTAB \cite{van1992bi}, MINRES \cite{paige1975solution}, or GMRES \cite{saad1986gmres} based on the Krylov subspace with appropriate preconditioners outperform \cite{langer2018direct} over the direct solver techniques in many complex problems. The computational cost can \textcolor{black}{be} of $O(N)$ or worse, and linear memory requirement $O(N)$ for the choice of the iterative solver and the preconditioner type \cite{bangerthLecture34}. The iterative solver does not solve the system exactly; it starts with an initial guess and continues to improve the solution in each iteration until the error/residual is less \textcolor{black}{than} a specified tolerance. The availability of good preconditioners is one of the main hurdles while using iterative solvers. Without good preconditioners, the issue of convergence arises, and the use of the iterative solver is not a good idea.

In this work, we \textcolor{black}{explore numerical solution methods to simulate the charging behavior of non-insulated superconducting pancake solenoids. The} mathematical model \textcolor{black}{considers} the magnetic field due to the current flow \textcolor{black}{in} complex \textcolor{black}{3D} geometries \textcolor{black}{in which a superconducting} wire \textcolor{black}{or coil} is co-wound with a conducting metal, rather than an insulator, separating the turns. \textcolor{black}{The superconductor is extremely thin with very high conductivity compared to the physical dimensions and conductivities of co-wound metals, and air. This introduces two extremely different time scales in the system which makes harder for the solvers to produce desire solutions.} The objective of this research work is to find efficient solvers for the present work that can scale to \textcolor{black}{multi-billions} degrees of freedom anticipated in future applications. The finite element solutions for the fully discrete scheme are obtained by using an iterative solver \textcolor{black}{combination (FGMRES-GMRES)} with AMS preconditioner \cite{hiptmair2006auxiliary, hiptmair2008auxiliary} which is built in the \textit{hypre} library \cite{falgout2002hypre}, and the direct solver MUMPS. \textcolor{black}{A} scalability analysis for both direct and iterative solvers \textcolor{black}{is} presented. \textcolor{black}{The scalability of the AMS preconditioner with conjugate-gradient (CG) method for low conductive materials is presented in \cite{kolev2009parallel}. To the best of our knowledge, the use of the robust FGMRES-GMRES solver with AMS preconditioner in simulating magnetic field with widely disparate material properties is new.} \textcolor{black}{All of the computational experiments presented in this paper are done in Petra-M \cite{Shiraiwa2017RF}.} 

The paper is organized as follows: In Section \ref{mathmodel} we present mathematical modeling for the governing equations of magnetic field simulation. \textcolor{black}{A brief description of the computational platform Petra-M is given in Section \ref{platform}.}
In Section \ref{numericalExp} we present the physical domains of the \textcolor{black}{non-insulated superconducting}  pancake solenoids \textcolor{black}{considered}, provide a details description of the numerical experiment techniques, represent the scalability analysis for both direct and iterative solvers. \textcolor{black}{Here the principal finding of this work is} shown in both tabular and graphical forms, that the iterative solver with AMS preconditioner outperforms over the direct solver \textcolor{black}{even with two extremely different time scales present in the system}. In Section \ref{conclusion-futuredir} we present conclusions and future research directions. \textcolor{black}{Finally, in the appendix, we show the Biot-Savart magnetic field computation for the spiral coils.}

\section{Mathematical Modeling}\label{mathmodel}
In this paper, we consider a time varying magnetic field which is excited by imposing voltage to a body of conducting coil. The conducting coil is not strictly contained in the computational domain $\Omega$ as its two disjoint electric ports $\Gamma_E$ and $\Gamma_J$ touch the domain boundary $\partial\Omega$. \textcolor{black}{We present the mathematical modeling of the time-dependent magnetic field following the frequency domain modeling in \cite{rodriguez2010eddy}.} We neglect the displacement current density because we are concern with diffusive and not with wave timescales. Therefore, the Ampere's law in differential form \cite{biro1999edge}: 
\begin{align}
    \nabla\times\bH=\bJ=\sigma\bE,
\end{align}
where $\bH$ the magnetic field strength, $\bJ$ the current density, $\bE$ the electric field strength, and $\sigma$ the conductivity. \textcolor{black}{For simplicity we consider $\sigma$ as a \textcolor{black}{positive} constant in this paper rather than a tensor.} Dividing both sides by $\sigma$, and taking \textit{curl} operator
\begin{align}
    \nabla\times\frac{1}{\sigma}\nabla\times\bH=\nabla\times\bE.\label{amp2}
\end{align}
Faraday's law of induction:
\begin{align}
    \nabla\times\bE=-\frac{\partial\bB}{\partial t}=-\mu_0\frac{\partial\bH}{\partial t},\label{Faraday}
\end{align}
for the magnetic flux density $\bB=\mu_0\bH$, where $\mu_0$ is a material independent parameter called the permeability constant. \textcolor{black}{The equation \eqref{Faraday} ensures that if the initial magnetic field strength is divergence free then $\nabla\cdot\bH=0$ holds for all time.} We assume \textcolor{black}{ $\Gamma_E$ and $\Gamma_J$} are perfect conductors, and thus the tangential component of the electric field vanishes there. Thus, we consider the following no-flux boundary conditions \cite{bossavit2000most, rodriguez2010eddy}
\begin{align}
    \bE\times\bnh=\frac{1}{\sigma}\nabla\times\bH\times\bnh&=\textbf{0}, \hspace{2mm}\text{in}\hspace{2mm}\Gamma_{E}\cup\Gamma_{J}\times(0,T],\label{portbc}\\
    \mu_0\bH\cdot\bnh&=0, \hspace{2mm}\text{in}\hspace{2mm}\partial\Omega\times(0,T], \label{nbc}
\end{align}
where $\bnh$ is the outward unit normal vector to the boundary, and $T$ the simulation time.  
\textcolor{black}{We apply a fixed non-zero voltage $V$ at $\Gamma_J$ and zero voltage at $\Gamma_E$, this potential difference drives the current to pass along the coil. Taking the dot product of $\bnh$ on both sides of \eqref{Faraday} and using \eqref{nbc} we have
    $\nabla\cdot(\bE\times\bnh)=0\hspace{2mm}\text{on}\hspace{2mm}\partial\Omega.$
}\textcolor{black}{Assuming the boundary $\partial\Omega$ a simply-connected surface, there exists a surface potential $v$ such that}
\begin{align}
    \bE\times\bnh=\frac{1}{\sigma}\nabla\times\bH\times\bnh=\nabla v\times\bnh\hspace{2mm}\text{on}\hspace{2mm}\partial\Omega,\label{bcs}
\end{align}
where $v_{|_{\Gamma_J}}=V$ \textcolor{black}{is a non-zero constant} and $v_{|_{\Gamma_E}}=0$.
Combining the equations \eqref{amp2}-\eqref{Faraday}, and \eqref{bcs}
we have the following time-dependent governing equations for the magnetic field simulation \textcolor{black}{of the voltage excitation problem}
\begin{align}
    \frac{\partial\bH}{\partial t}+\frac{1}{\mu_0}\nabla\times\frac{1}{\sigma}\nabla\times\bH &=\textbf{0},\hspace{10mm}\forall(x, t)\in\Omega\times(0,T],\label{gov1}\\
    \textcolor{black}{\frac{1}{\sigma}\nabla\times\bH\times\bnh} &=\textcolor{black}{\nabla v\times\bnh},\hspace{1mm}\forall(x,t)\in\partial\Omega\times(0,T],\label{gov2}\\
    \bH(x,0) &=\textbf{0}, \hspace{11mm}\forall x\in\Omega,\label{gov4}
\end{align}
where $v_{|_{\Gamma_J}}=V$ and $v_{|_{\Gamma_E}}=0$. 

\textcolor{black}{We note that this is a diffusion equation for the magnetic field with characteristic time given by $t_{\rm d} = \mu_0\sigma l^2$, \textcolor{black}{where $l$ is a characteristic length scale}. In the simulations shown in the following sections, the time is normalized by this diffusion time, where $l = 1.14m$ and $\sigma = 2\times 10^{6}S/m$ are used. Additionally, the magnetic fields are normalized by the field produced by a single turn coil with same current ($I =$1000A), defined as $B_{0}:= \mu_0I/l$. }



\section{Software and Computational Facility}\label{platform}
\begin{figure}[ht!]
    \begin{center}
\includegraphics[width=1\textwidth,height=0.25\textwidth] {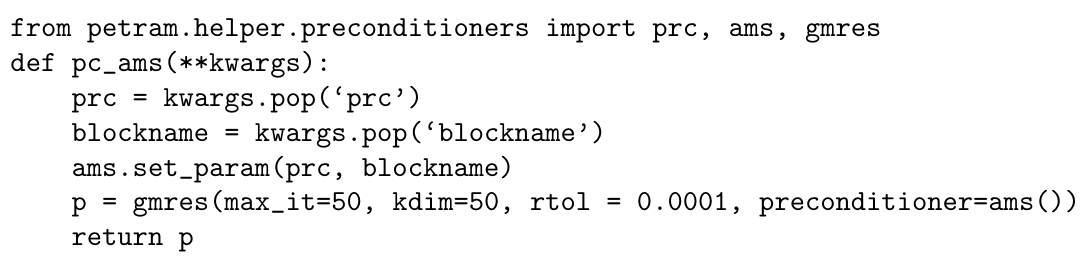}
\caption{\textcolor{black}{Petra-M code segment defines preconditioned inner solver.}}\label{petram_code}
\end{center}
\end{figure}

The time-dependent equations \eqref{gov1}-\eqref{gov4} was discretized fully by the backward-Euler timestepping scheme and solved on the Petra-M \cite{Shiraiwa2017RF} finite element analysis platform. 
This open source platform allows for constructing a geometry, creating a mesh, assembling and solving the finite element linear system, and solving and visualizing the results using a user friendly graphical interface (GUI). 
Petra-M uses various open source software. In particular, it uses MFEM modular finite element library \cite{mfem-library} for the FEM linear system assembly. 
The weakform PDE interface in Petra-M allows for defining a mixed form PDE system we solved in this paper by choosing form integrators available in the MFEM library from menus.
Petra-M combines GUIs with the python scripting, allowing for rapidly developing the simulation model with very little coding effort. 
For example, the inner solver with pre-conditioner is defined in the code segment is given Fig. \ref{petram_code}. 
Then, the outer solver is configured to use this inner solver using GUI.
All simulations are done \textcolor{black}{on} the `Engaging' cluster computer at \textcolor{black}{the} Massachusetts Institute of Technology, in which one node consists of 32 cores and 512 GB RAM \textcolor{black}{memory}.


\section{Numerical Experiments}
\label{numericalExp}
For our simulations, we consider four different geometries for the charging pancake solenoids: One turn single pancake (T-1), ten turns double pancakes (T-10) with each pancake has five turns, twenty turns double pancakes (T-20) with each pancake has ten turns, and thirty turns double pancakes (T-30) with each pancake has fifteen turns. The cross-section of the HTS spiral coils are rectangular. \textcolor{black}{The thickness and height of the HTS coil are $0.16cm$ and $0.3cm$, respectively for all of the four models.} In each of the geometry, the HTS coil is \textcolor{black}{co-wound with}  copper with a rectangular cross-section, \textcolor{black}{which is then co-wound with stainless steel. The copper and stainless steel co-wounds  help not to arise numerical issues as they prevent the superconducting coil from directly touching the air.} The copper \textcolor{black}{co-wound} \textcolor{black}{spiral} coil preserves the spiral pattern \textcolor{black}{of} the HTS \textcolor{black}{coil}, keeping the uniform gap between turn to turn. \textcolor{black}{A sufficiently large air domain is considered surrounding the stainless steel body so that the magnetic field lines do not go beyond it.} Finally the terminals of the copper \textcolor{black}{co-wound} HTS coil are extended \textcolor{black}{to the air boundary} by solid copper rectangular bars. \textcolor{black}{The two copper terminals are extended to the boundary so that external voltage can be applied to one of them, and the voltage drives the current to flow along the spiral coil and the magnetic field is produced. Thus, in this work, we consider forced voltage excitation coil \cite{meng2020effect,rodriguez2010eddy}, and the voltage is adjusted to have the same current for all four models \cite{by201416}. }

The physical properties of the different models are described below:
\subsection{T-1 Model}

\begin{figure}
    \centering
    \begin{minipage}{0.5\textwidth}
        \centering
        \includegraphics[width=0.3\textwidth, height=0.2\textwidth, viewport=200 0 1050 800]{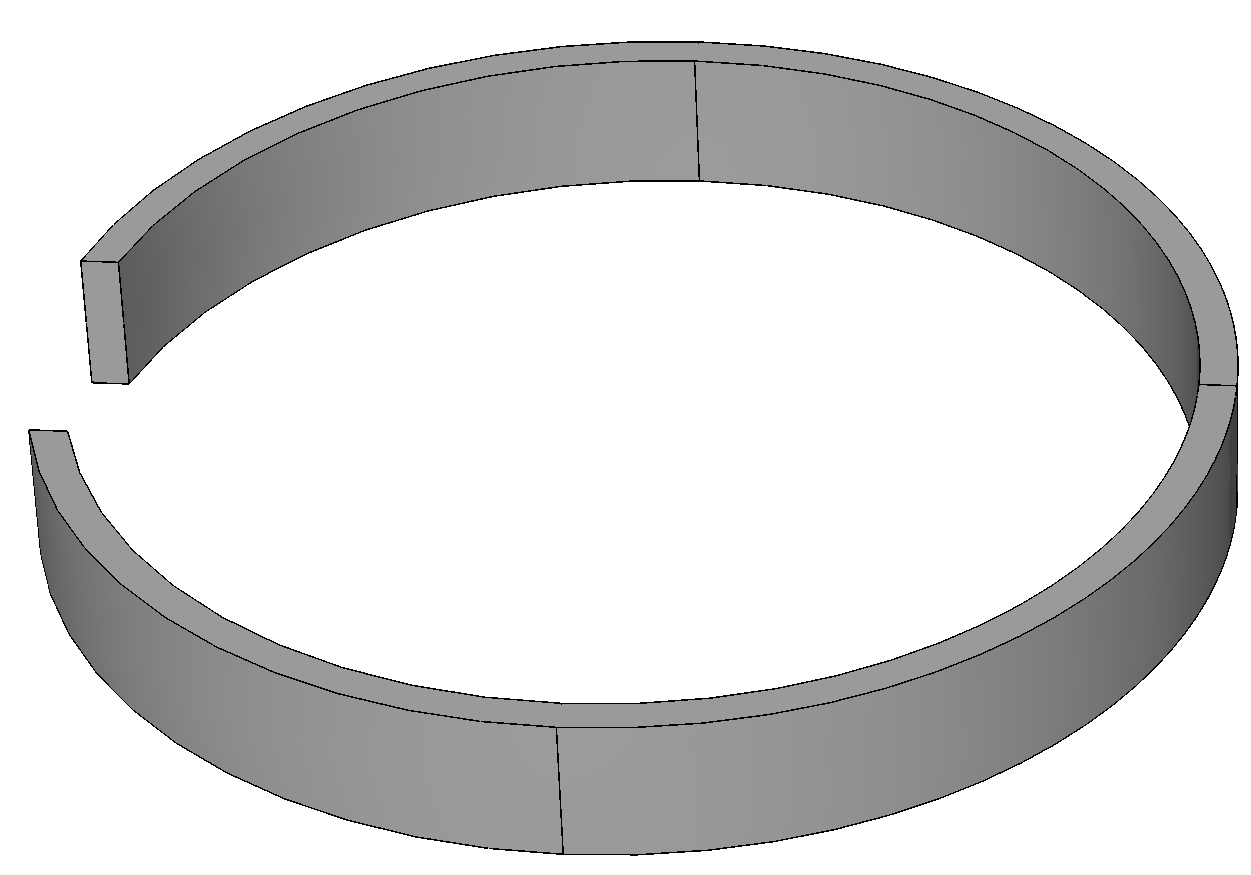} 
    \end{minipage}\hfill
    \begin{minipage}{0.5\textwidth}
        \centering
        \includegraphics[width=0.4\textwidth,height=0.4\textwidth, viewport=200 0 1050 950]{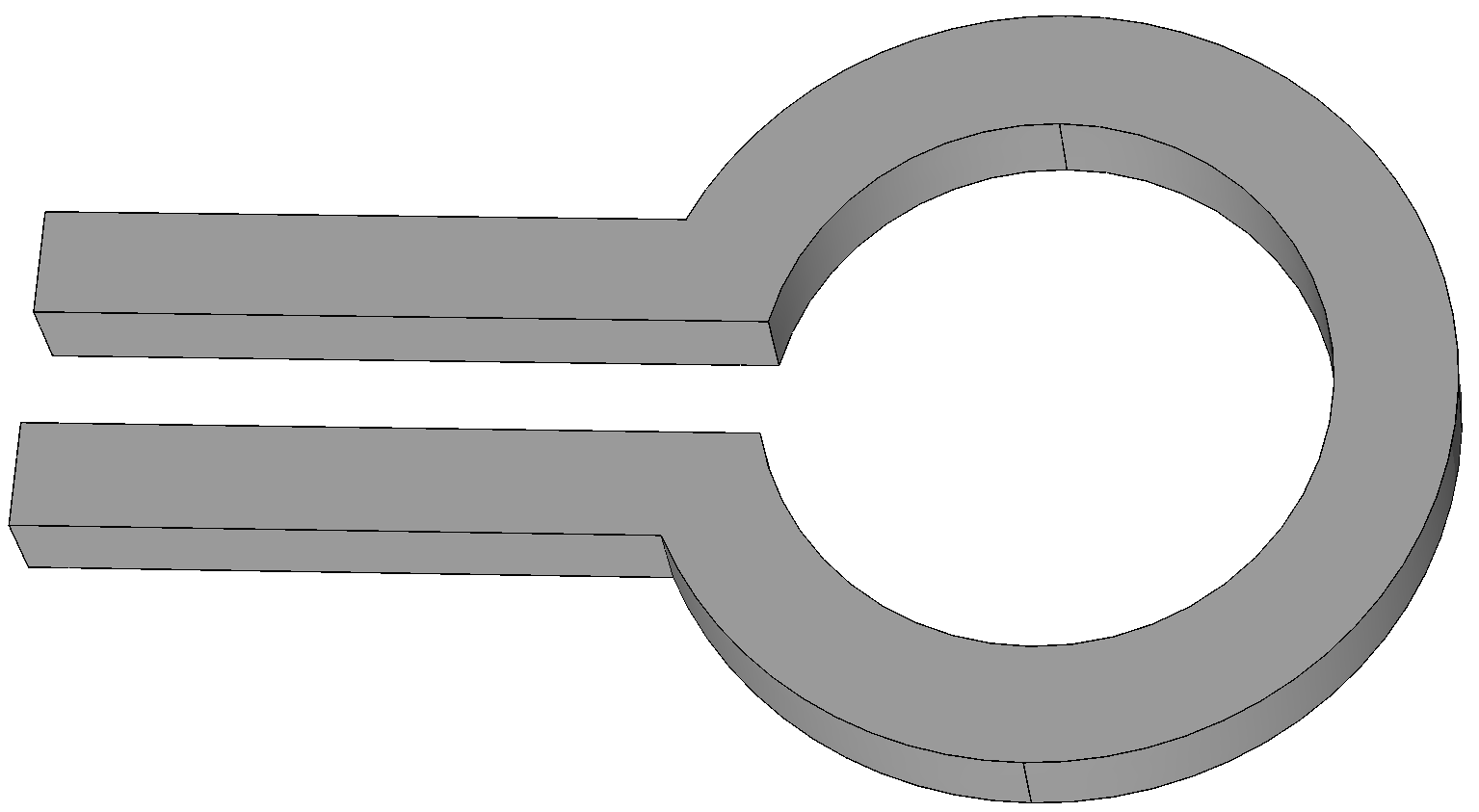} 
    \end{minipage}
    \caption{HTS \textcolor{black}{turn} (left) and copper \textcolor{black}{co-wound} coil (right) \textcolor{black}{with electric ports} in \textcolor{black}{T-1} model.}\label{T-1model-hts-copper}
\end{figure}

\begin{figure}[ht!]
    \begin{center}
\includegraphics[width=0.5\textwidth,height=0.35\textwidth] {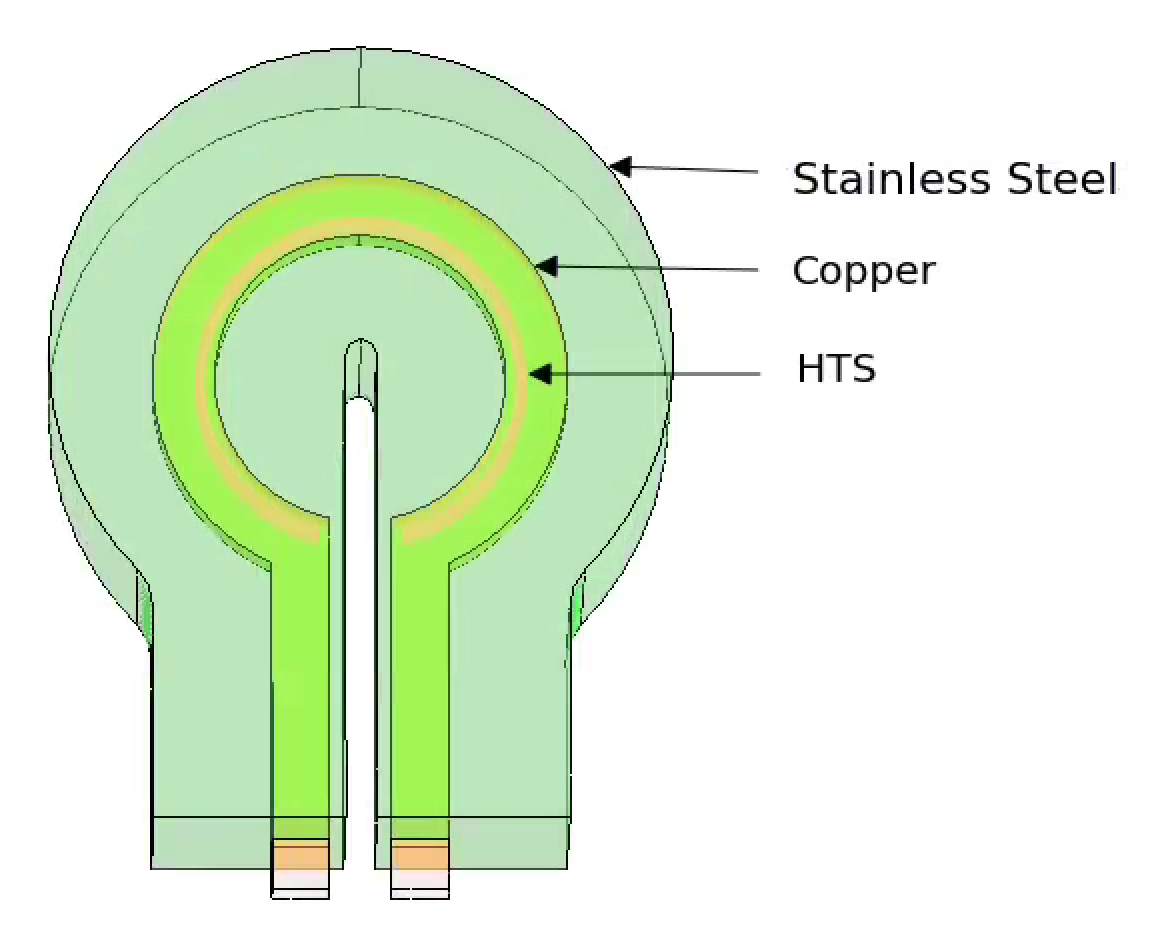}
\caption{\textcolor{black}{Stainless steel co-wound of the copper co-wound coil in T-1 model.}}\label{T-1model-ss-air}
\end{center}
\end{figure}

In the single pancake T-1 model solenoid, the HTS \textcolor{black}{coil} has only one turn \textcolor{black}{with center at the origin}.  A $1 m^3$ box \textcolor{black}{centered at the origin,} is considered as the air boundary. The Figures \ref{T-1model-hts-copper}-\ref{T-1model-ss-air} show the shape of different material components in the T-1 model. 

\subsection{T-10 Model}
\begin{figure}
    \centering
    \begin{minipage}{0.5\textwidth}
        \centering
        \includegraphics[width=0.4\textwidth, height=0.6\textwidth, viewport=200 0 1100 400]{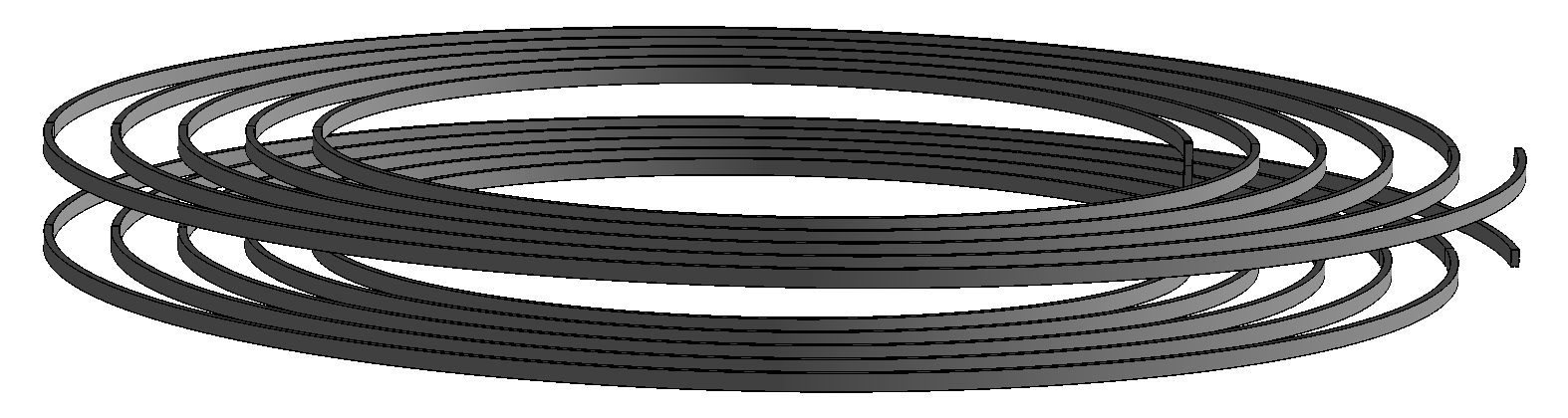} 
    \end{minipage}\hfill
    \begin{minipage}{0.5\textwidth}
        \centering
        \includegraphics[width=0.4\textwidth,height=0.25\textwidth, viewport=200 0 1050 600]{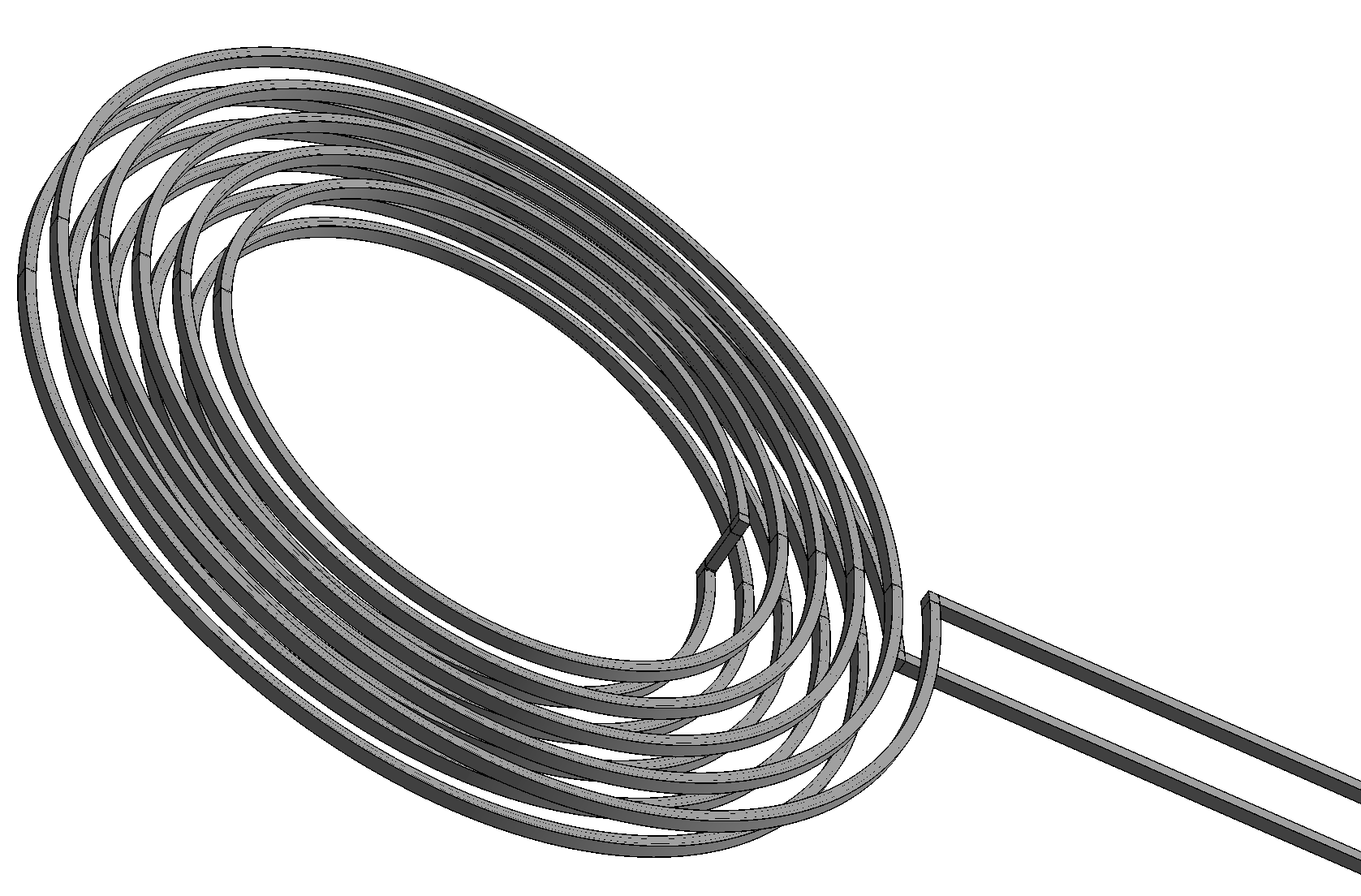} 
    \end{minipage}
    \caption{HTS \textcolor{black}{coil} (left) and copper \textcolor{black}{co-wound} coil \textcolor{black}{with electric ports} (right) in \textcolor{black}{T-10} model.}\label{T-10model-hts-copper}
\end{figure}

\begin{figure}[ht!]
    \begin{center}
\includegraphics[width=0.5\textwidth,height=0.25\textwidth] {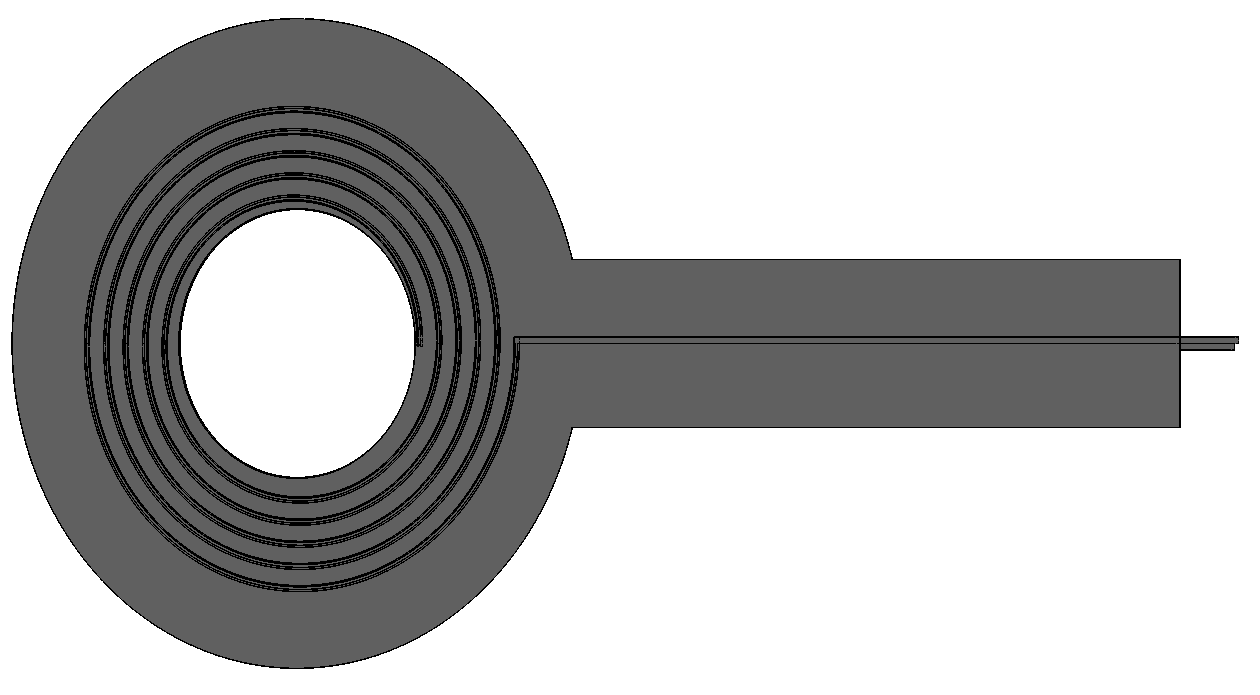}\\
\includegraphics[width=0.5\textwidth,height=0.08\textwidth] {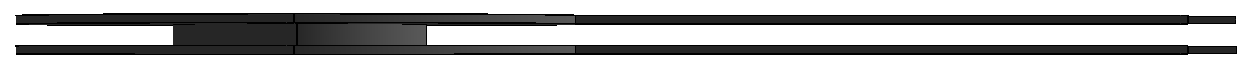}
\caption{Top and lateral views of stainless steel boundary in \textcolor{black}{T-10} model.}\label{T-10model-ss}
\end{center}
\end{figure}

The T-10 model geometry is a double pancake solenoid. Each of the pancakes has five HTS turns. The turns in the upper \textcolor{black}{pancake spiral in and the turns in the lower pancake spiral out so that the current direction remains same}, and their innermost turns are connected by a vertical joggle. \textcolor{black}{The gap between two turns is 1.32 cm.}  \textcolor{black}{The HTS coil and its copper co-wound with two electric ports are shown in Fig. \ref{T-10model-hts-copper}. The electric ports are extended to the air boundary.} The copper co-wound HTS is again co-wound with stainless keeping the air gap between two pancakes as shown in Figure \ref{T-10model-ss}. \textcolor{black}{For the finite element simulations, we consider a cylindrical air domain of base diameter $100cm$, height $114cm$, and its axis is parallel to the pancakes so that the magnetic field lines remain inside the computational domain.}

\subsection{T-20 Model} 
\begin{figure}[ht!]
    \begin{center}
\includegraphics[width=0.5\textwidth,height=0.25\textwidth] {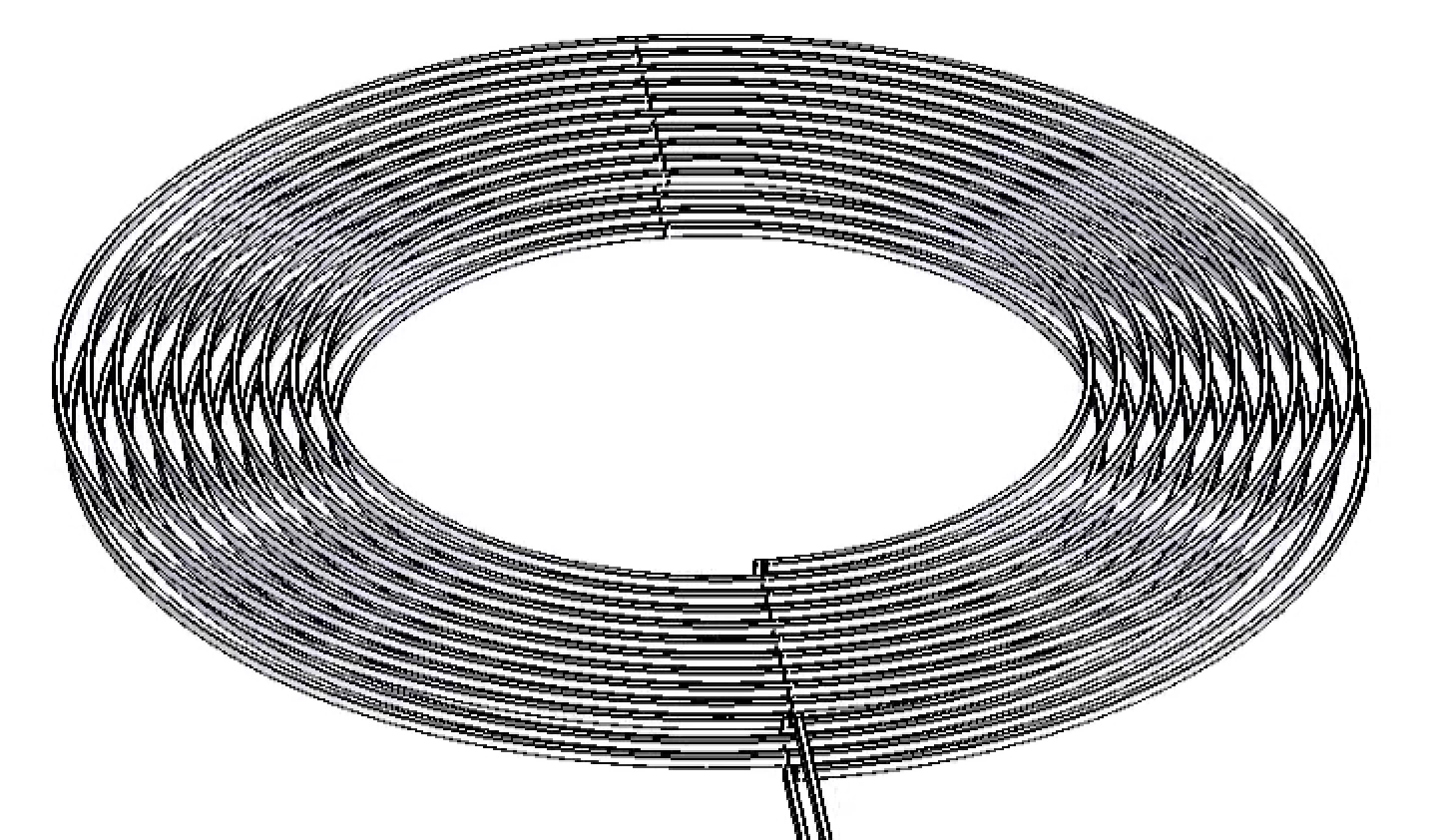}
\caption{HTS \textcolor{black}{coil with electric ports} in T-20 model.}\label{T-20model-hts}
\end{center}
\end{figure}

The T-20 model geometry is also a double pancake solenoid and its physical properties are similar to the T-10 model except each of the pancakes has ten turns HTS spiral coil. \textcolor{black}{The gap between two turns is 0.65 cm.} \textcolor{black}{The HTS coil with two electric ports of the T-20 model is showing in Fig. \ref{T-20model-hts}.} \textcolor{black}{The physical properties of the air domain are the same as those in the T-10 model.} 

\subsection{T-30 Model} 
\begin{figure}
    \centering
    \begin{minipage}{0.5\textwidth}
        \centering
        \includegraphics[width=0.3\textwidth, height=0.35\textwidth, viewport=300 0 800 1350]{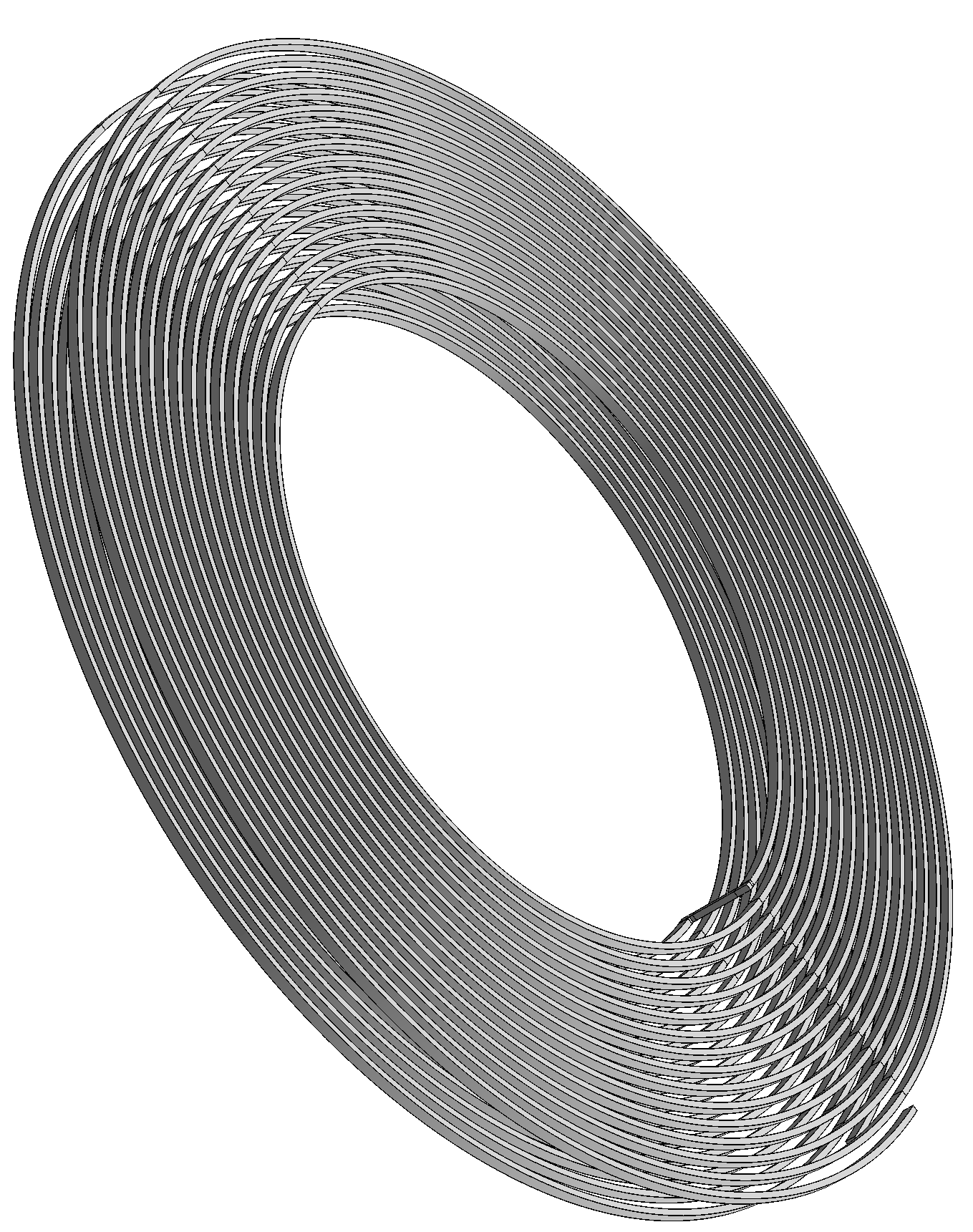} 
    \end{minipage}\hfill
    \begin{minipage}{0.5\textwidth}
        \centering
        \includegraphics[width=0.3\textwidth,height=0.6\textwidth, viewport=300 0 1500 1100]{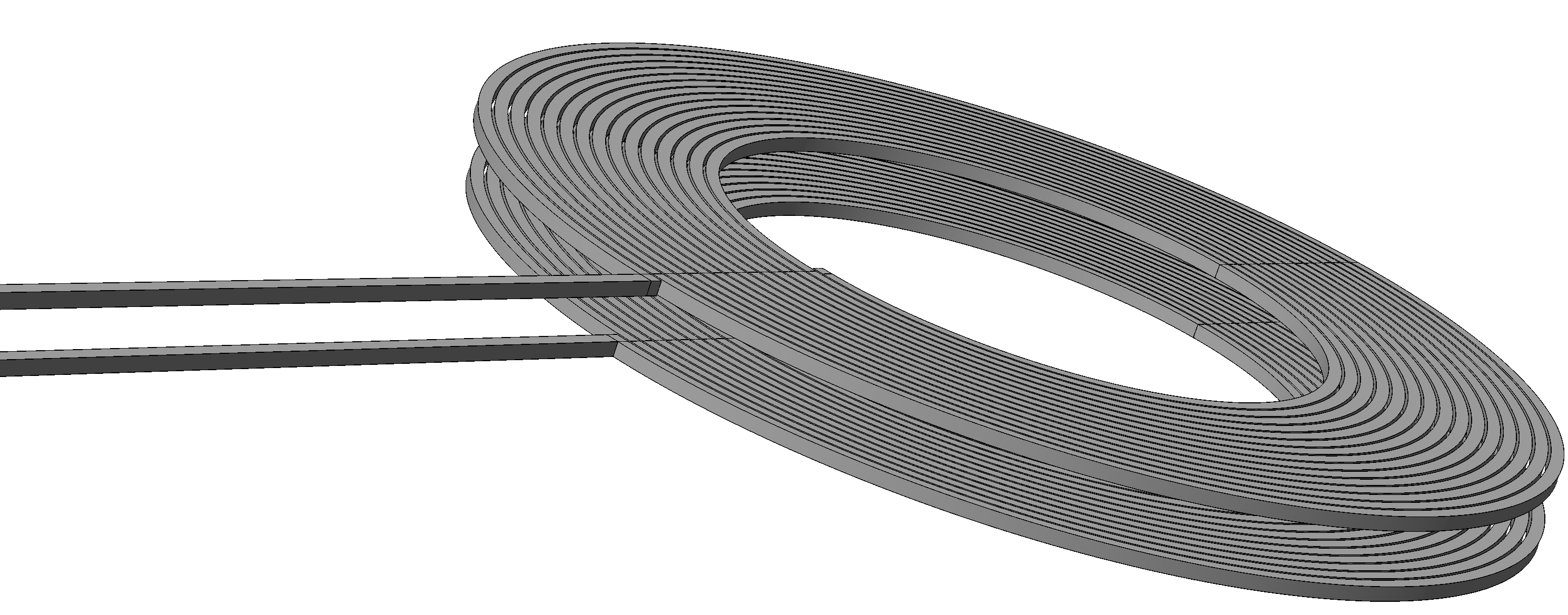} 
    \end{minipage}
    \caption{HTS \textcolor{black}{coil} (left) and copper \textcolor{black}{co-wound HTS} \textcolor{black}{coil with two electric ports} (right) in T-30 model.}\label{T-30-model-hts-copper}
\end{figure}

The T-30 model is a double pancake charging solenoid where each pancake has a fifteen turns HTS coil. Dimensions of each component of the T-30 model are kept almost the same as those in the T-10 and T-20 models. \textcolor{black}{The gap between two turns is 0.45 cm.} The HTS coil and copper \textcolor{black}{co-wound} HTS \textcolor{black}{coil} with electric ports in the T-30 model are showing in Figure \ref{T-30-model-hts-copper}. \textcolor{black}{The physical properties of the air domain are also the same as those in T-10 model.}

\subsection{Iterative Method}
For the iterative solver, we consider the flexible inner-outer Krylov subspace methods \cite{saad1993flexible,simoncini2003flexible} which allow varying the preconditioner from one iteration to another. Using Krylov methods, instead of solving the linear system in equation \eqref{system1}, we solve a modified system such as the following right preconditioned equation
\begin{align}
    AM^{-1}M\bx=\bb,
\end{align}
into two steps as
\begin{align}
    AM^{-1}\by=\bb, \hspace{2mm}\text{with}\hspace{2mm}M\bx=\by,
\end{align}
having a fixed preconditioner $M$. In flexible inner-outer method, we are allowed to use a different matrix, say $M_j$, at each iteration. In inner-outer approach, the preconditioner itself can be a Krylov subspace method, and thus very appealing. 

FGMRES \cite{saad1993flexible} is used as the outer solver and GMRES as the inner solver. FGMRES stands for flexible GMRES, which is a variant of GMRES method but more robust. FGMRES algorithm is essentially the GMRES algorithm but with variable preconditioning. Except for the variable preconditioning, the only difference from the standard GMRES is that in FGMRES the preconditioned vectors are saved and used them to update the solution.  For constant preconditioner, the FGMRES and GMRES algorithms are mathematically equivalent. In this paper, we use the AMS preconditioner to the GMRES solver and GMRES solver preconditioner to the FGMRES solver. 

\subsection{Iterative Solver Selection} In this experiment, we show the performance of FGMRES, GMRES, and BiCGSTAB as inner or outer solver with AMS and GS (Gauss-Seidel) preconditioners.  We consider T-1 model with $6137426$ \textcolor{black}{dofs}, simulation end time $T=0.5$, time step size $\Delta t=0.5$ (that is, a single time step solve), 4 nodes, 64 cores, and the parameters in the inner-outer solvers are given in Table \ref{parameter_set}.
\begin{table}[ht!]
\begin{center}
	\begin{tabular}{|c|c|c|c|c|c|c|}\hline
		\multicolumn{3}{|c|}{Inner Solver} & \multicolumn{4}{|c|}{Outer Solver}\\\hline
		Max It.   & restart & rtol  & Max It. & rtol. & abs. tol. & restart\\\hline
		50        & 50   & $10^{-4}$  & 2000 & $10^{-8}$ & $10^{-10}$ & 100\\\hline
		\end{tabular}\caption{Parameters those are used in the flexible inner-outer method.}\label{parameter_set}
	\end{center}
	\end{table}
	
We record the number of iterations and solving time in Table \ref{solver-select}, taken by a combination of solvers and preconditioners. We observe, FGMRES as the outer solver and GMRES as the inner solver with AMS preconditioner takes $4$ iterations with least solving time $31.43s$, and thus choose this combination for all simulations with the iterative solver.
\begin{table}[ht!]
	\begin{center}
		\begin{tabular}{|c|c|c|c|c|}
			\hline
			  Outer Solver & Inner Solver   & Preconditioner  & Iteration & Time      \\
			\hline
		
			 FGMRES &  GMRES & GS & Does not converge &-- \\\hline
			 FGMRES &  BiCGSTAB & AMS &$3$ &$44.30s$ \\\hline
			 FGMRES &  GMRES & AMS &$4$ & 31.43s \\\hline
             FGMRES &  FGMRES & AMS & $4$&  40.09s\\\hline
             GMRES  &  FGMRES & AMS & 5 & 49.34s\\\hline
             GMRES &  GMRES & AMS &$4$ & $34.21s$ \\\hline
		\end{tabular}\caption{Performance of different solvers in the Inner-Outer method.}\label{solver-select}
	\end{center}
\end{table}

As a direct solver, we use MUMPS. The lowest order edge element is used for all simulations. The scalability analysis is performed with the iterative and direct solvers for T-1, T-10, T-20, and T-30 models. We used the following parameters values:
$\sigma_{hts}=1.0\times 10^{15}S/m$,
$\sigma_{cu}=4.01\times 10^8 S/m$, $\sigma_{fe}=2.0\times 10^6S/m$, and  $\sigma_{air}=1.0\hspace{1mm}S/m$, as the conductivity of HTS, copper, stainless steel, and air, respectively, and the magnetic permeability constant $\mu_0=4\pi\times 10^{-7} H/m$.

\begin{figure}[ht!]
    \begin{center}
\includegraphics[width=.7\textwidth,height=0.40\textwidth]{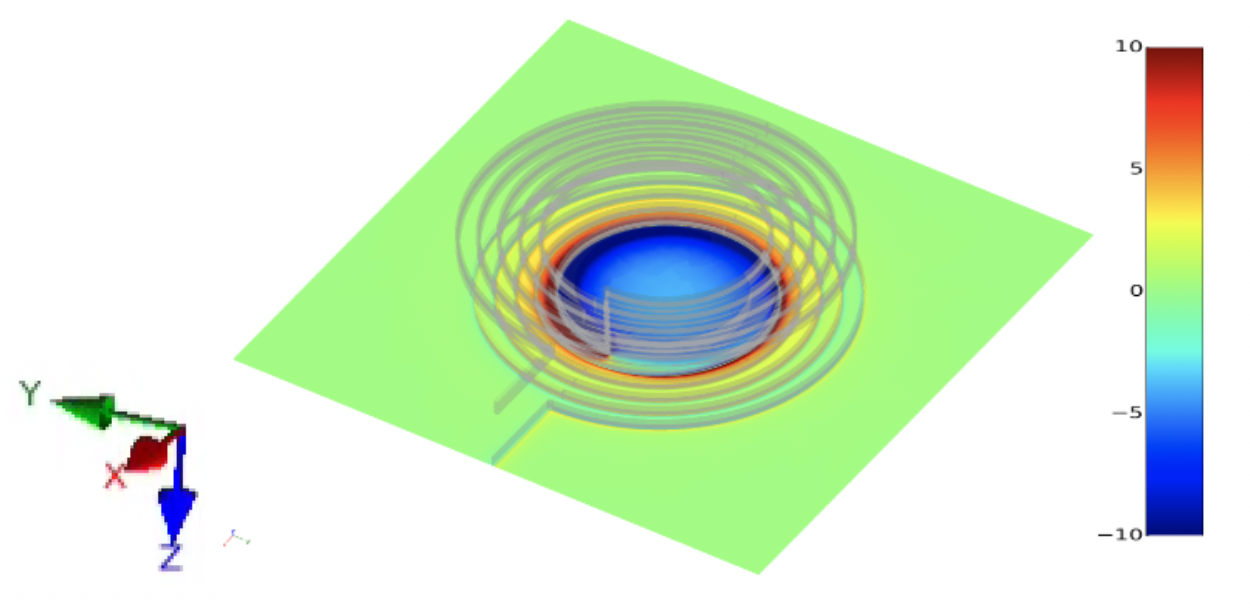}
\caption{Plot of the cross section along the $xy$-plane of $\bB_z$ in T-10 model (lower view).}
\end{center}
\end{figure}

\subsection{Code Verification and Model Comparison} 

To verify our finite element method (FEM) code and to compare the four different models, we compute $\bB$ at the origin. We compare the stationary (fully charged) phase of $\bB$ with the magnetic field computed by the Biot-Savart law. We also compare the magnetic field profiles for all four models.

The voltage excitation problem is solved using FEM so that the same current $I=1000A$ remains on the ports for all the models. We used coarse meshes of the domains, a normalized timestep size $\Delta t=3.06$, and ran the simulations until the normalized end time $T=460$ for each of the models.  For all these simulations, we used the direct solver.


The analytical formula for $\bB$ due to the current flow in the spiral \textcolor{black}{HTS coil} is derived by \textcolor{black}{assuming} the  \textcolor{black}{coil as} filament, \textcolor{black}{representing it} as a vector equation and using the Biot-Savart law. In the Biot-Savart law, the uniform current $I=1000A$ is assumed everywhere in the filament. \textcolor{black}{The Biot-Savart computation of the magnetic field for the spiral coils are presented in Appendix A.}

In Table \ref{b_max_center}, we present the FEM magnetic field at the stationary phase and the magnetic field computed from the Biot-Savart law. We observe a good agreement among the results of the two methods. \textcolor{black}{Assuming the Biot-Savart magnetic fields as a benchmark, we computed the relative errors of the FEM magnetic fields and presented them in the last row. All the relative errors are less than $3\%$.} The tabular values clearly show that as the number of turns increases the field strength gets stronger. \textcolor{black}{Moreover, it is observed that the field strengths in T-20, and T-30 models are approximately double and triple as found in the T-10 model, which seems consistent as the field strength should be proportional to the number of turns. Since in T-1 model, the HTS does not have a full turn, we found a weaker field strength than what it supposed to be for a full one-turn model.}


\begin{table}[ht!]
	\begin{center}
		\begin{tabular}{|c|c|c|c|c|}\hline
	     \backslashbox{Method}{Model} &T-1 & T-10 & T-20 & T-30\\\hline
Biot-Savart   & 3.02 & 49.71 &   98.07 & 148.60\\\hline
FEM  &  2.94   & 50.98  & 99.25 & 150.14\\\hline
\textcolor{black}{Relative Error} &\textcolor{black}{2.63\%} &\textcolor{black}{2.55\%} &\textcolor{black}{1.20\%}&\textcolor{black}{1.04\%}  \\\hline
		\end{tabular}\caption{For each model, the fully charged magnetic field strength at the origin.}\label{b_max_center}
	\end{center}
\end{table}

\begin{figure}[ht!]
    \begin{center}
\includegraphics[width=0.9\textwidth,height=0.40\textwidth]{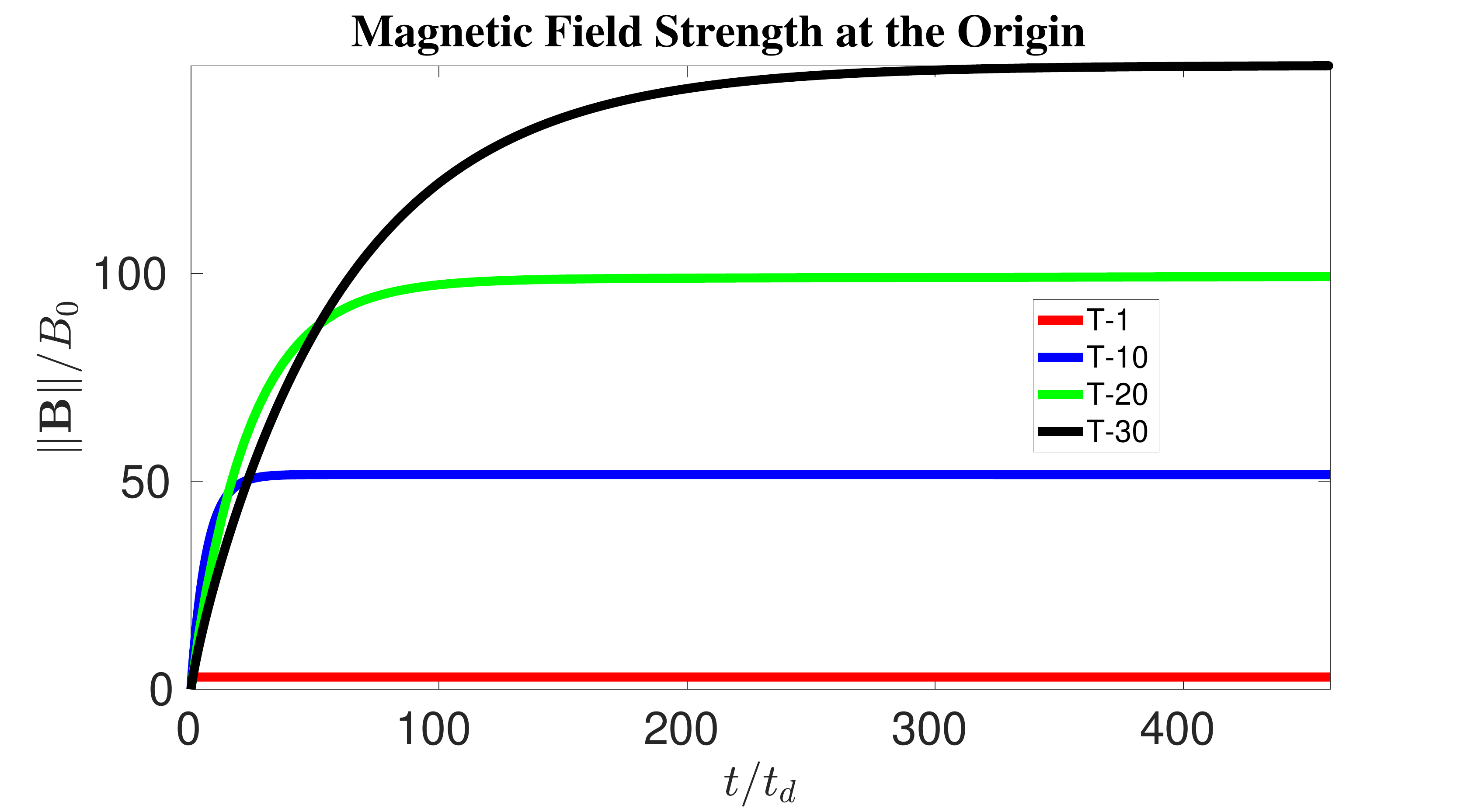}
\caption{Charging time for T-1, T-10, T-20, and T-30 models.}\label{profile_b0}
\end{center}
\end{figure}

For each of the model, the magnetic field profile is computed, normalized and presented altogether in Fig. \ref{profile_b0}. These profiles give us the full charging time of the models and observed that as the number of turns increases, the charging time increases as well. \textcolor{black}{That is, the more turns in the coil the model has higher inductance.}

\subsection{Number of turns versus computational time}

Now we want to compare the four different models in terms of the required computational time \textcolor{black}{by the iterative solver} for a single timestep solve in the FEM simulations. Since the inductances of the models are not the same, for a fair comparison, first we determine the appropriate timestep size for \textcolor{black}{each of} them. From the computed fully charged value of $\|\bB\|$ at the origin, we determine its $90\%$ growth time, and divide it by a fixed number and set it as the timestep size $\Delta t$ of a model. We use the fixed number equal to 100 in this experiment. We find \textcolor{black}{$\Delta t=0.0098, 0.1439, 0.5572$, and    $1.4083$} for the T-1, T-10, T-20, and T-30 model, respectively. \textcolor{black}{We generate meshes for each of the models so that they all provide closely to 2.6 million dofs.} With the above stated timestep sizes, we run the simulations of the respective model, use the proposed iterative solver with a single core processor to avoid communication time and record the single timestep solving time in Table \ref{single_solve_time}.  From the fourth column in Table \ref{single_solve_time}, we see that as the number of turns increases the solving time increases. As the number of turns increases, the length of the HTS coil becomes longer, and consequently, it's contribution to the system matrix gets stronger. Thus, the system matrix becomes more ill-conditioned and it becomes harder to solve for the iterative solver.

\textcolor{black}{We note that for all these four simulations we do the profiling of our code, and find that the sparse matrix-vector multiplication in the iterative process is a major time consuming step.}

\begin{table}[ht!]
	\begin{center}
		\begin{tabular}{|c|c|c|c|}\hline
	Model      &dofs & $\Delta t$ & Total Wall Clock Time \\\hline
T-1 & 2751768           & 0.0098     & 418.75s  \\\hline
T-10  & 2664168&  0.1439& 8860.03s\\\hline
	T-20  & 2680759           & 0.5572     & 10108.69s \\\hline 
T-30  & 2624896 & 1.4083& 23990.02s \\\hline
		\end{tabular}\caption{\textcolor{black}{Time step used in simulation and solving time required by the iterative solver for four models. Note that time step is chosen to be 1\% of the charging time. }}\label{single_solve_time}
	\end{center}
\end{table}

\subsection{Computational time versus conductivity}

\begin{figure}[ht!]
    \begin{center}
\includegraphics[width=0.8\textwidth,height=0.450\textwidth]{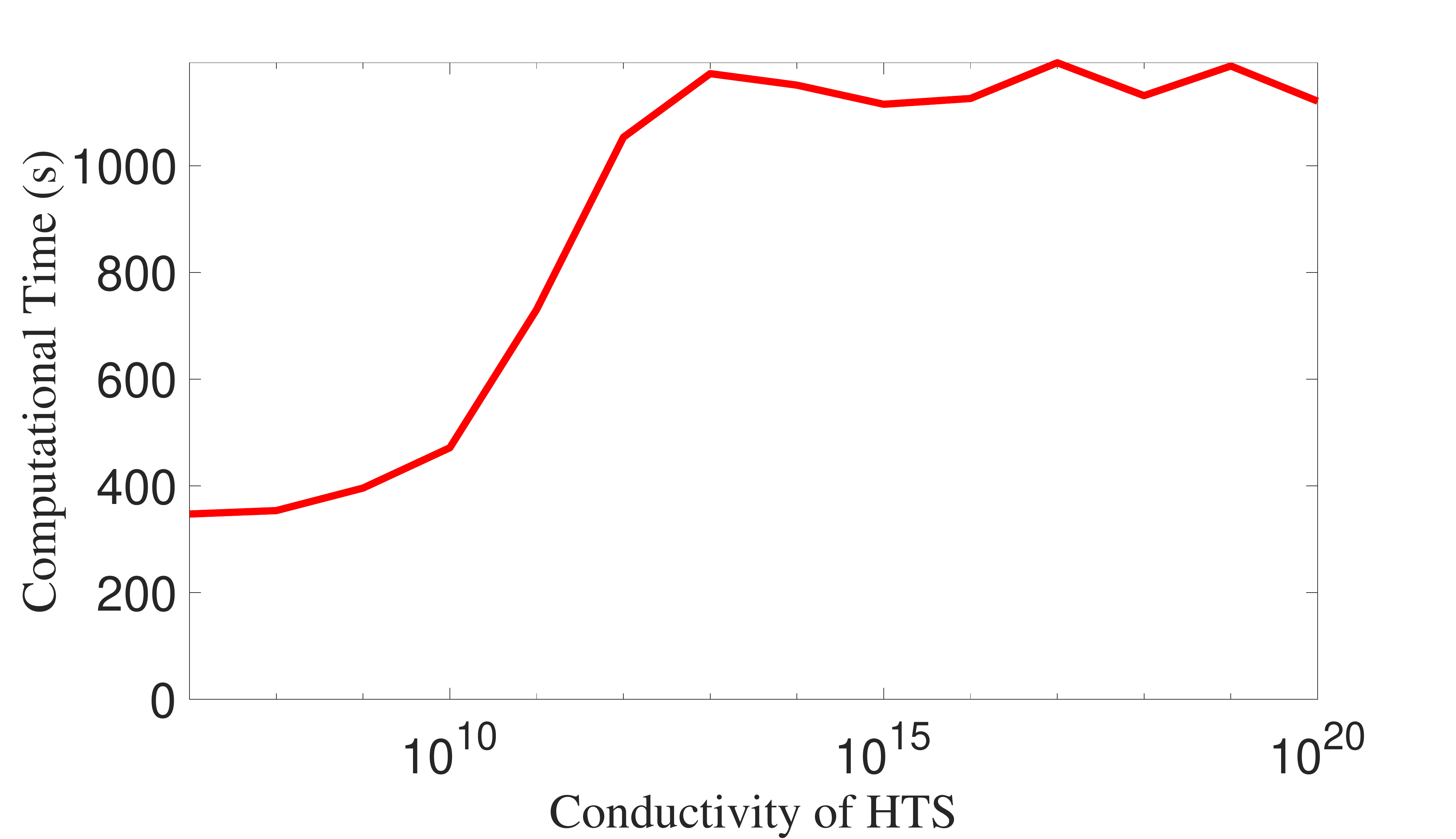}
\caption{Computational time of the iterative solver versus the conductivity of HTS.}\label{conductivity}
\end{center}
\end{figure}

In this section, we investigate how the computational time of the iterative solver varies as the \textcolor{black}{assumed} conductivity of the HTS increases. We consider the T-30 model with a problem size of 330963 \textcolor{black}{dofs}, use 1 node with 1 core, and normalized time stepsize $\Delta t=1.4083$. \textcolor{black}{We vary the conductivity from $10^7$ to $10^{20}$ uniformly and run the simulations for a single time step solve, that is, $T=1.4083$.} The \textcolor{black}{computational times} are recorded in Table \ref{conductivity-time} \textcolor{black}{and represented in Fig. \ref{conductivity}. We observe that the solving time increases sharply as the conductivity of the HTS increases from $10^{7}$ to $10^{13}$. \textcolor{black}{If we increase the conductivity further, the computational time remains almost of the same order.} As the assumed conductivity of HTS increases, it starts introducing two extremely different time scale in the system, and thus the system matrix becomes more ill-conditioned and consequently, it becomes harder for the iterative solver to solve the system.}  


\begin{table}[ht!]
\begin{center}
	\scriptsize
		\begin{tabular}{|c|c|c|c|c|c|c|c|}\hline
		 $\sigma_{hts}(S/m)$&
		 $10^{7}$ &$10^{8}$&$10^{9}$&$10^{10}$&$10^{11}$&$10^{12}$&$10^{13}$ \\\hline
		Wall Clock Time (s) & 347.25 & 353.67   &  395.91 & 471.44 & 730.44 &    1053.17    & 1172.45\\\hline\hline
		 $\sigma_{hts}(S/m)$ &$10^{14}$&$10^{15}$&$10^{16}$&$10^{17}$&$10^{18}$&1e19&$10^{20}$ \\\hline
		 Wall Clock Time (s) & 1150.90   &  1115.03 & 1125.89   & 1192.85 &    1131.21& 1186.79 & 1120.84\\\hline
		\end{tabular}\caption{Solving time \textcolor{black}{increases as the conductivity of HTS does.}}\label{conductivity-time}
	\end{center}
\end{table}

\subsection{Parallel Scalability Analysis}
Parallel scalability analysis is widely used \cite{he2020efficient, huo2020designing,ono2020scalable,shoukourian2014predicting, sun1994scalability} to measure the performance of parallel codes as the problem size and the number of computer processor cores increase. It helps to predict the performance of a large number of cores on a large problem size based on the performance of a small number of cores on small problem size. In this section, we investigate the parallel scalability for the two solvers, the direct solver MUMPS, and the preconditioned iterative solver FGMRES-GMRES in terms of the weak scaling, weak scaling efficiency, \textcolor{black}{speed up, and strong scaling efficiency}.

\subsubsection{Weak Scaling} If we increase the number of cores in such a way that even the problem size increases but the workload on each core remains the same, then it refers as weak scaling \cite{shoukourian2014predicting}. 
\subsubsection{Weak Scaling Efficiency (WSE)} If the amount of time to complete a work unit on one unit core(s) is $C_1$, and the amount of time to complete the same work of $p$ units on $p$ units cores is $C_p$, then the weak scaling efficiency is defined as $WSE:=\frac{C_1}{C_p}*100\%$.
\subsubsection{Weak Scaling Results and Discussions}
All the recorded \textcolor{black}{total wall clock} time herein is for a single time step solving time with normalized timestep size $\Delta t=0.15$.
\begin{table}[ht!]
\begin{center}
		\begin{tabular}{|c|c|c|c|c|c|c|}\hline
		Nodes  & Cores      &dofs& MUMPS & WSE &FGMRES-GMRES & WSE\\\hline
		1   & 2   & 88437  &  14.35s &     &  7.77s &  \\\hline
		1   & 16    & 687425 & 77.03s &  18.63\%    & 16.00s & 48.56\%\\\hline
		4(5)  & 128    & 5415890  & 787.48s &1.82\% & 43.38s & 17.91\%\\\hline
		\end{tabular}\caption{Weak scaling:  T-1 model.} \label{t-1WS}
	\end{center}
\end{table}

\begin{figure}
    \centering
    \begin{minipage}{0.45\textwidth}
        \centering
        \includegraphics[width=1\textwidth,height=0.75\textwidth]{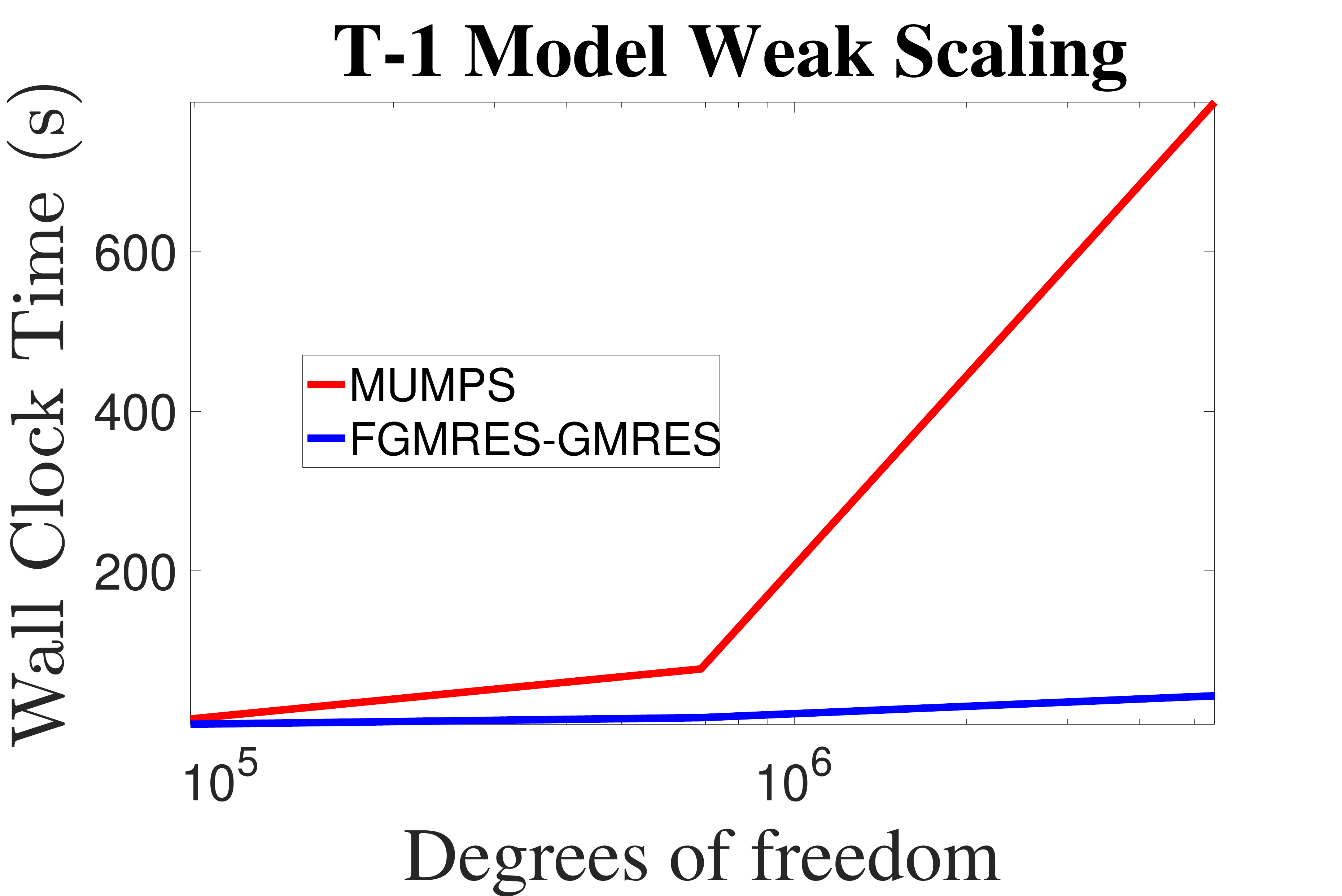}
        \caption{Measured computational time versus degrees of freedom.}\label{t-1ws-time}
    \end{minipage}\hfill
    \begin{minipage}{0.45\textwidth}
        \centering
        \includegraphics[width=1\textwidth,height=0.75\textwidth]{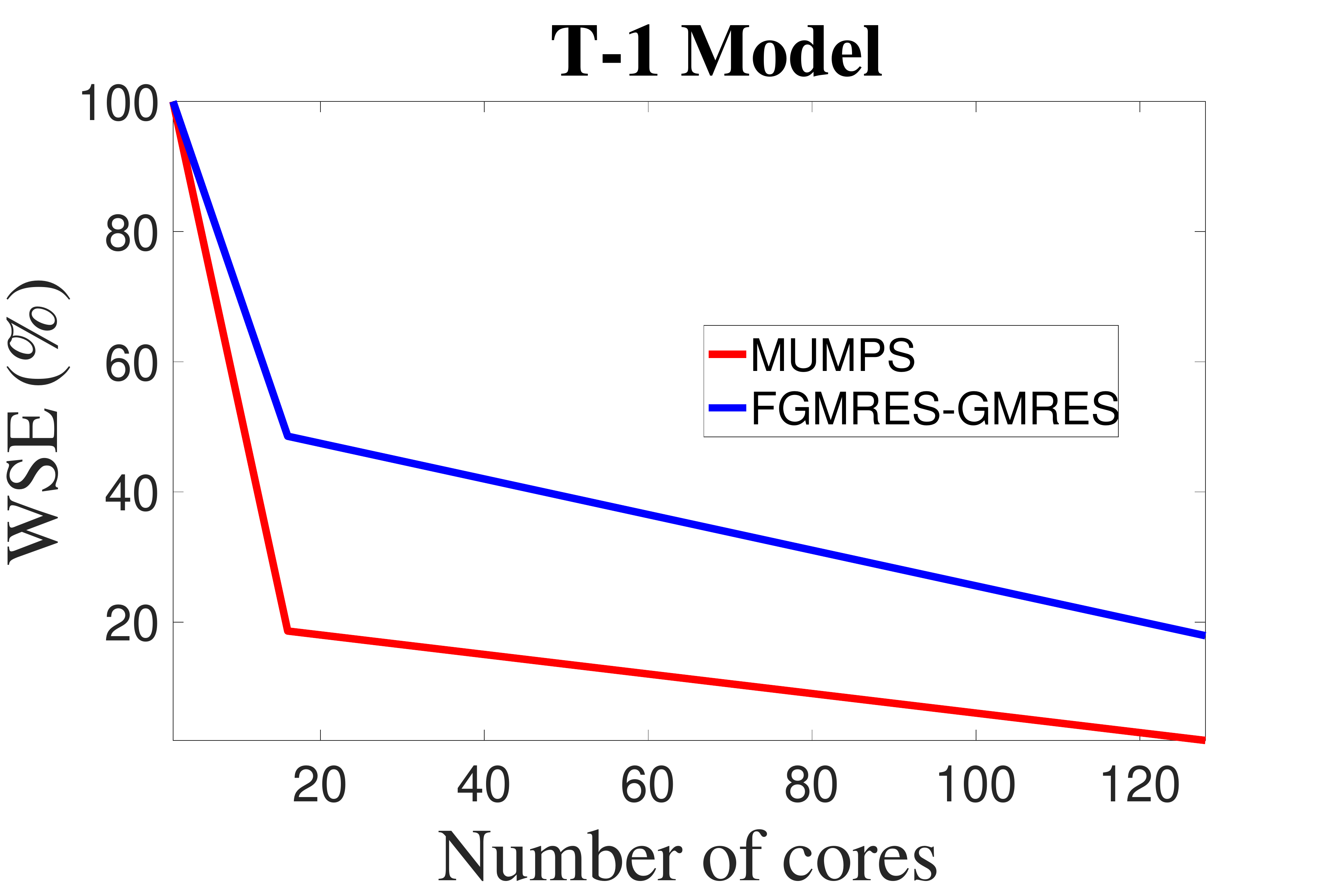} 
        \caption{WSE versus number of cores employed.}\label{t-1wse-cores}
    \end{minipage}
\end{figure}
Table \ref{t-1WS} shows the weak scaling performances of the direct and iterative solvers with the T-1 model. At first, we solve the problem of size 88437 dofs with 1 node and 2 cores. In the next refinement, the problem size is of 687425 dofs which is approximately 8 times the previous problem's dofs. To keep the workload the same on each core, we increase the number of cores to 16, solve the problem, and record the solving time. We repeated the same procedure in the next refinement when the problem size increased to 5415890 dofs. In this case, the direct solver could not solve the problem without using less than 5 nodes while the iterative solver used only 4 nodes, that is, the iterative solver required less memory than the direct solver. \textcolor{black}{The computational time taken by the solvers to solve a problem was represented as wall clock time.} We plotted the \textcolor{black}{wall clock time}  versus the dofs for both the direct and iterative solvers in Fig. \ref{t-1ws-time}. It is observed that the direct solver computational time remains always higher than the computational time of the iterative solver, that is, there is no crossing point. Also, as the problem size increases the computational time of the direct solver becomes much higher than that of the iterative solver.

The weak scaling efficiency is calculated as defined above and plotted against the number of cores employed in Fig. \ref{t-1wse-cores}. We observe the weak scaling efficiency of the iterative solver is higher than that of the direct solver. As the problem size increases, the direct solver weak scaling efficiency drops to a factor of 10 while the iterative solver weak scaling efficiency drops only to a factor of 3.   


\begin{table}[ht!]
\begin{center}
		\begin{tabular}{|c|c|c|c|c|c|c|}\hline
		Nodes  & Cores      &dofs& MUMPS & WSE &FGMRES-GMRES & WSE\\\hline
		1   & 5   & 273837  &  25.77s &     &  93.14s &  \\\hline
		1&16 & 853921 & 56.07s&45.96\% &130.80s&71.21\% \\\hline
		2 & 34  & 1860729 & 83.82s &30.74\% & 164.82s&56.51\% \\\hline
		2 & 40&   2168158 &89.38s &28.83\%& 221.17s& 42.11\%  \\\hline
		4& 124& 6771172 & 275.80s& 9.34\%& 512.87s& 18.16\% \\\hline
		10&316& 17289872& 1002.37s&2.57\%&576.32s& 16.16\% \\\hline
	\end{tabular}\caption{Weak scaling:  T-10 model.}\label{t-10WS}
	\end{center}
\end{table}

\begin{figure}
    \centering
    \begin{minipage}{0.45\textwidth}
        \centering
        \includegraphics[width=1.\textwidth,height=0.75\textwidth]{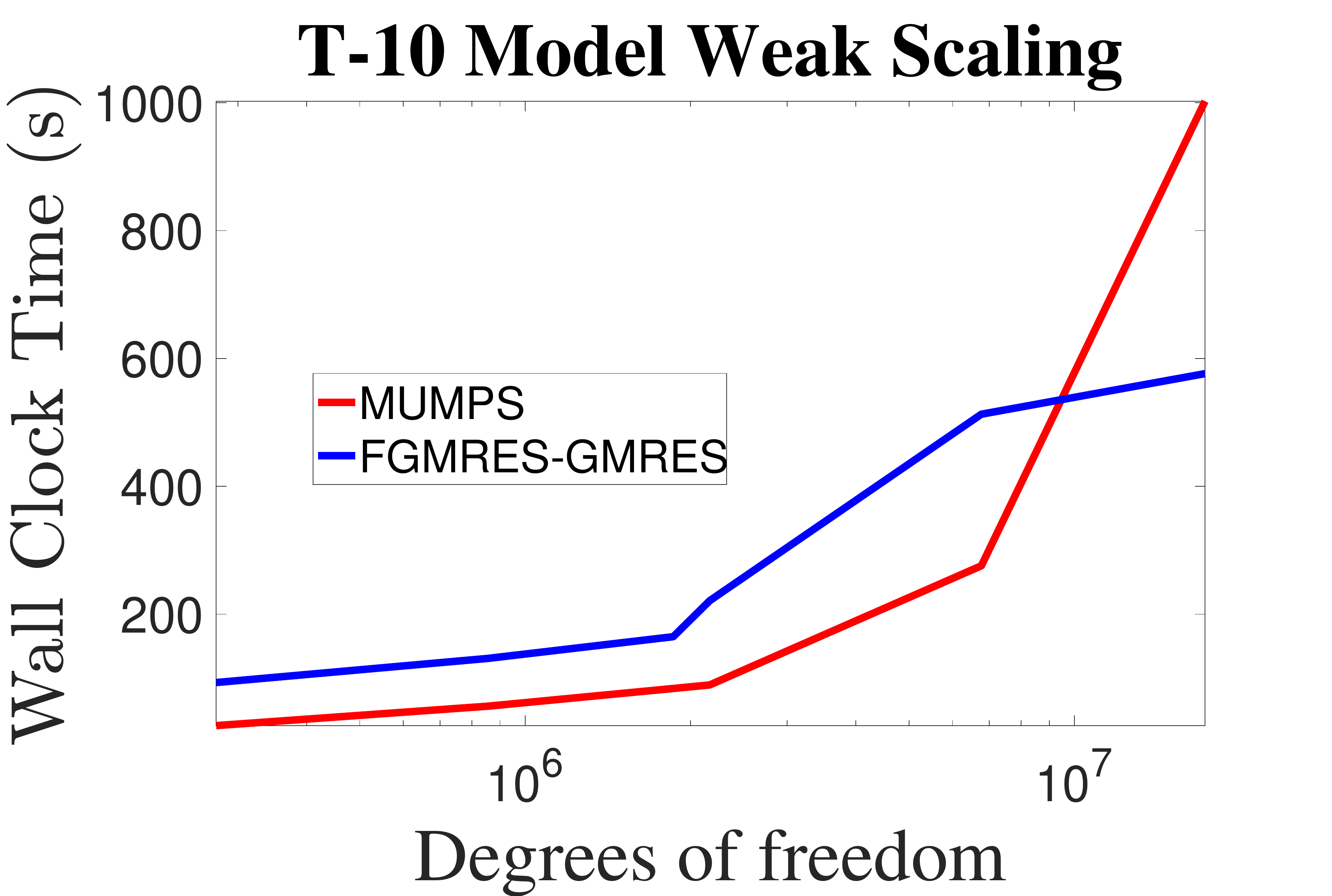} 
        \caption{Measured computational time versus \textcolor{black}{degrees of freedom.}}\label{t-10wse-time-dof}
    \end{minipage}\hfill
    \begin{minipage}{0.45\textwidth}
        \centering
        \includegraphics[width=1.\textwidth,height=0.75\textwidth]{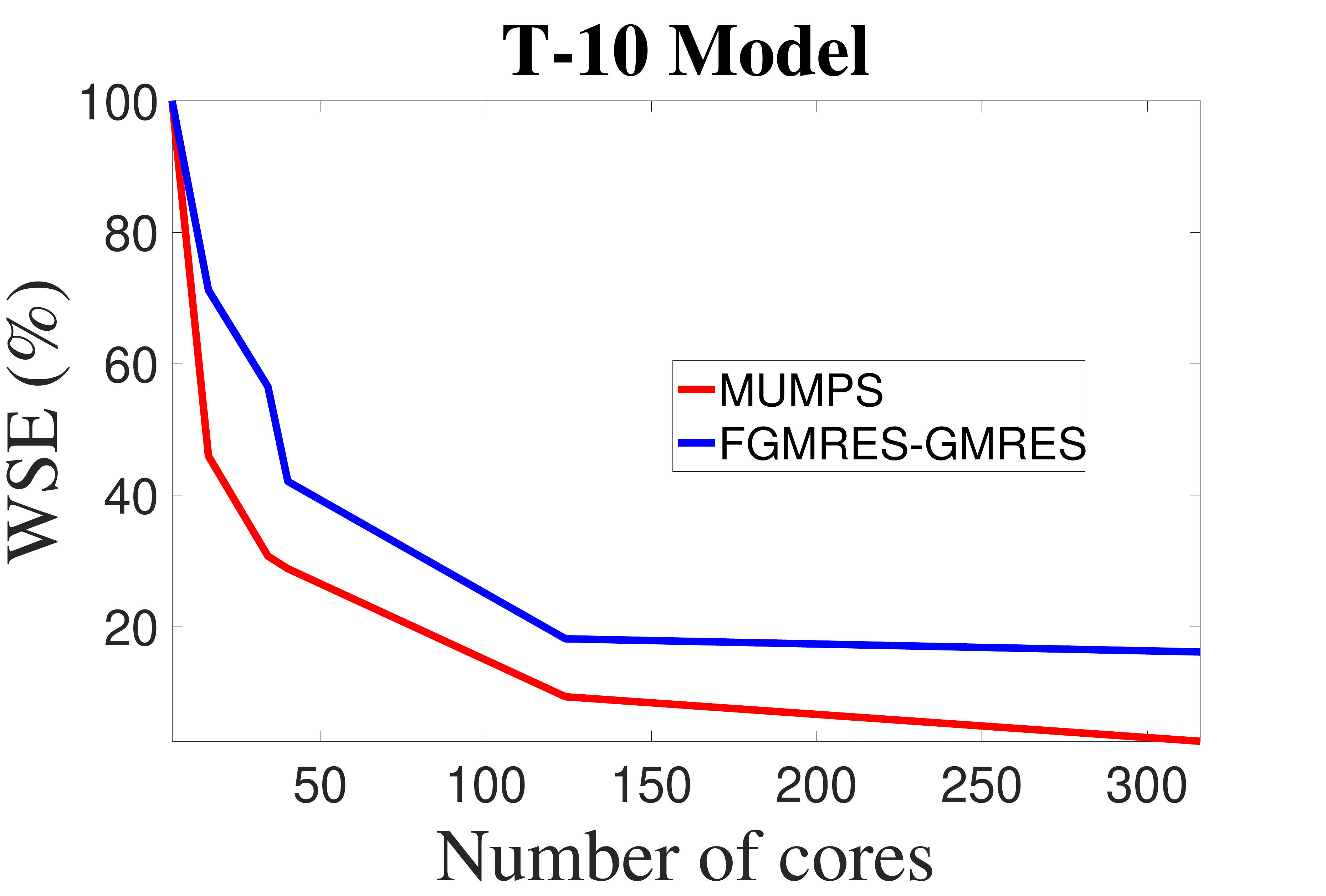} 
        \caption{WSE versus number of cores employed.}\label{t-10wse-cores}
    \end{minipage}
\end{figure}

To study the weak scaling of the solvers with the T-10 model, we solve the problem with 273837, 853921, 1860729, 2168158, 6771172, and 17289872 \textcolor{black}{dofs} employing 5, 16, 34, 40, 124, and 316 cores, respectively so that the workload on each core remains same. We recorded the computational time and WSE for both solvers and represented in Table \ref{t-10WS}. We plotted the wall clock time versus dofs in Fig. \ref{t-10wse-time-dof}. We notice that for problems with lower \textcolor{black}{dofs}, the direct solver outperforms over the iterative solver, but for higher \textcolor{black}{dofs} the preconditioned iterative solver solves the problem much faster than the direct solver. And at around 10 million dofs there is a crossing point where the direct solver computational time exceeds the iterative solver time.

We also plotted the WSE versus the number of cores employed in Fig. \ref{t-10wse-cores}. We observe that as we increase the number of cores employed, the iterative solver WSE remains always higher than that of the direct solver.

\begin{table}[ht!]
	\begin{center}
		\begin{tabular}{|c|c|c|c|c|c|c|}\hline
		Nodes  & Cores    &dofs& MUMPS & WSE &FGMRES-GMRES & WSE\\\hline
		1   & 5  & 1094858 &  186.39s &     &  301.85s &   \\\hline
		2   & 25    & 5418938 & 446.72s &  41.72\%    & 550.87s & 54.80\%\\\hline
		8(12)  & 86    & 18628074 & 1175.38s & 15.86\% & 730.84s & 41.30\%\\\hline
		\end{tabular}\caption{Weak scaling:  T-20 model.}\label{t-20WS}
	\end{center}
\end{table}
\begin{figure}
    \centering
    \begin{minipage}{0.45\textwidth}
        \centering
        \includegraphics[width=1.0\textwidth,height=0.75\textwidth]{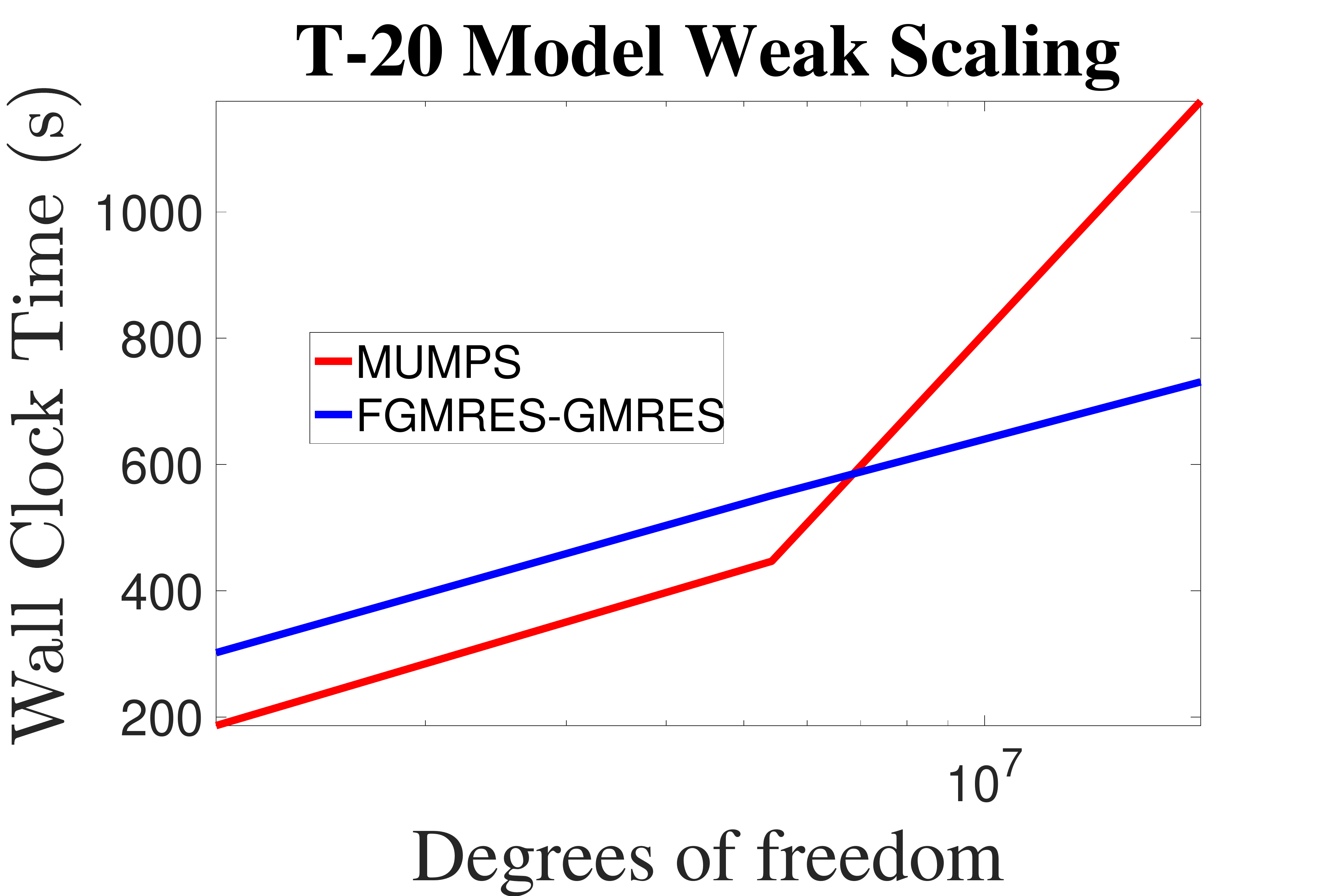} 
        \caption{Measured computational time versus \textcolor{black}{degrees of freedom.}}\label{t-20wse-time-dof}
    \end{minipage}\hfill
    \begin{minipage}{0.45\textwidth}
        \centering
        \includegraphics[width=1.0\textwidth,height=0.75\textwidth]{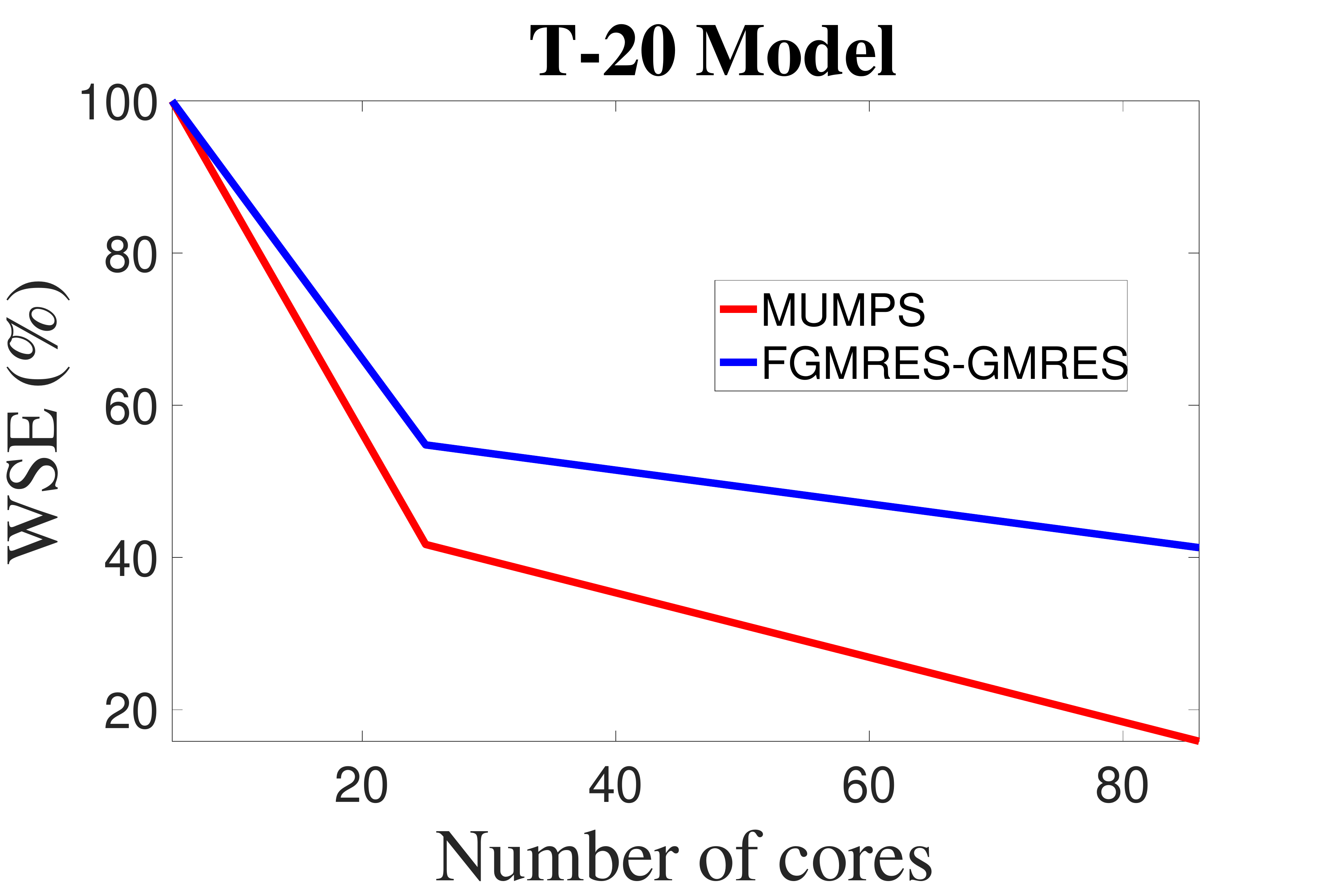} 
        \caption{WSE versus number of cores employed.}\label{t-20wse-cores}
    \end{minipage}
\end{figure}

Table \ref{t-20WS} represents the weak scaling performance with the T-20 model. In this case, the initial problem was a size of 1094858 dofs, and we solved it with 1 node and 5 cores. In the next refinement, the \textcolor{black}{dofs} increases to 5418938 which is 5 times more than before, and thus we solve this using 2 nodes and 25 cores. We repeat this process one more time and record the solving time. That is $5418938/1094858\approx 25/5$ and $18628074/5418938\approx 86/25$. When the problem gets 18628074 dofs, the minimum memory requirement for the direct solver is 12 nodes while the iterative solver can solve it using only 8 nodes. We plotted the solving time versus \textcolor{black}{dofs} in Fig. \ref{t-20wse-time-dof}. It shows, for low \textcolor{black}{dofs}, the computational time taken by the direct solver is less than that of the iterative solver. But as the problem complexity in terms of the \textcolor{black}{dofs} increases, the direct solver computational time becomes much higher than that of the iterative solver. We observed that the direct solver wall clock time exceeds the iterative solver wall clock time at around 6 million dofs.

From Fig. \ref{t-20wse-cores}, we also observe, the weak scaling efficiency of the iterative solver is much higher than that of the direct solver. As the \textcolor{black}{dofs} increases the direct solver weak scaling efficiency drops at a higher rate compared to the iterative solver.

\begin{figure}
    \centering
    \begin{minipage}{0.45\textwidth}
        \centering
        \includegraphics[width=1.0\textwidth,height=0.75\textwidth]{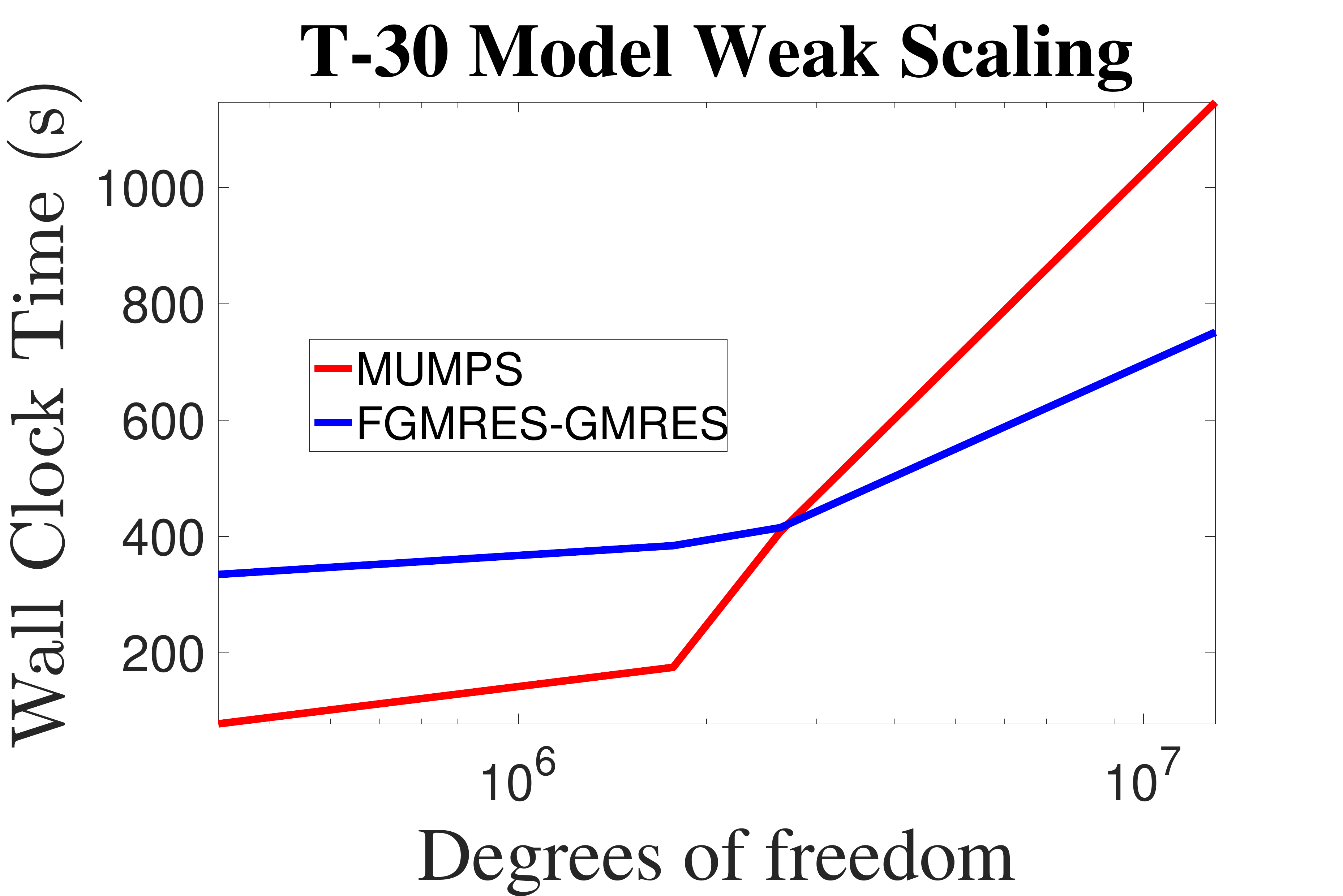} 
        \caption{Measured computational time versus \textcolor{black}{degrees of freedom}.}\label{t-30wse-time-dof}
    \end{minipage}\hfill
    \begin{minipage}{0.45\textwidth}
        \centering
        \includegraphics[width=1.0\textwidth,height=0.75\textwidth]{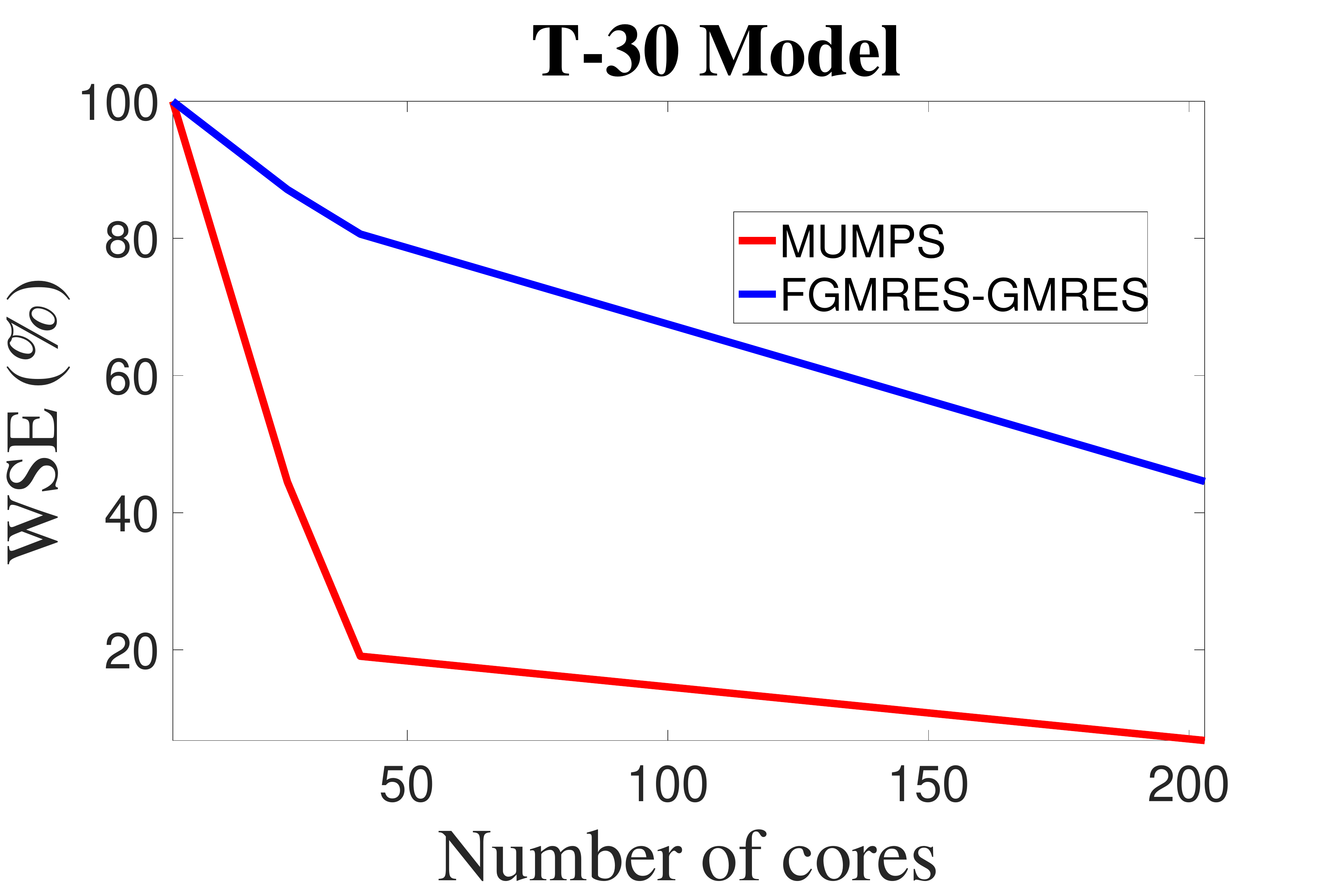} 
        \caption{WSE versus number of cores employed.}\label{t-30wse-cores}
    \end{minipage}
\end{figure}

\begin{table}[ht!]
\begin{center}
		\begin{tabular}{|c|c|c|c|c|c|c|}\hline
		Nodes  & Cores    &dofs& MUMPS & WSE &FGMRES-GMRES & WSE\\\hline
		1   & 5   & 330963  &  77.89s &     &  334.74s &  \\\hline
		1  & 27      & 1768029  & 175.02s &  44.50\%    & 384.18s & 87.13\%\\\hline
		 2  & 41  & 2624896 & 408.32s & 19.08\%  & 415.15s & 80.63\%\\\hline
  		 8 & 203 & 13031432 & 1146.46s & 6.79\% &751.22s & 44.56\%\\\hline

		\end{tabular}\caption{Weak scaling:  T-30 model.}\label{t-30WS}
	\end{center}
\end{table}
We also studied the weak scaling with the T-30 model and presented the results in Table \ref{t-30WS}. In this case, we consider 5 cores as unit cores and solve the problem with a coarse mesh of 330963 \textcolor{black}{dofs} using both direct and iterative solver. We observe the direct solver is faster than the iterative solver. For the finer meshes as the number of \textcolor{black}{dofs} increases, we increase the number of cores so that the workload on each core remains the same. Which is done as keeping the ratios in the degrees of freedom the same as the ratios in the number of cores, i.e. $1768029/330963\approx 27/5$, $2624896/330963\approx 41/5$, and $13031432/330963\approx 203/5$. We plotted the solving time versus \textcolor{black}{dofs} in Fig. \ref{t-30wse-time-dof}, and weak scaling efficiency versus the number of cores in Fig. \ref{t-30wse-cores}. We observe that for coarse mesh the direct solver outperform over the iterative solver but as the mesh becomes finer and consequently the \textcolor{black}{dofs} increases the computational time of the direct solver increases faster than that of the iterative solver. Also, there is a crossing point on the graph, which is at around 3 million dofs, after that the direct solver time graph overshoots the iterative solver time graph. The weak scaling efficiency of the iterative solver always lies above the direct solver weak scaling efficiency.

Finally, from Fig.  \ref{t-10wse-cores}, \ref{t-20wse-time-dof}, and \ref{t-30wse-time-dof}, we observe that as the number of turns increases the crossing point (where the direct solver computational time exceeds the iterative solver time) abscissa  reduces. That is, if the number of turns of a coil increases, even the problem size remains small, the preconditioned iterative solver wins over the direct solver.

\subsubsection{Strong Scaling} 
We study the strong scalability performance of the direct and iterative solvers by keeping fixed the problem size in terms of the \textcolor{black}{dofs} and increasing the number of cores used to solve it.
\subsubsection{Speedup ($S_p$) and Strong Scaling Efficiency (SSE)} We use the most commonly used metric, \textit{speedup}, for the strong scaling analysis \cite{sun1994scalability}. Among several types of \textit{speedup} metric, we use the \textit{fixed-size speedup}, where the problem size remains fixed and determine how faster the problem can be solved. If the computational time to solve a problem of size $N$ with one unit core(s) is $C_1(N)$, and the computational time to solve the same problem with $p$ units cores is $C_p(N)$, then $S_p$ and 
SSE are defined as $S_p:=\frac{C_1(N)}{C_p(N)}$ and $SSE:=\frac{C_1(N)}{p\hspace{1mm}C_p(N)} *100\%$, respectively. In an ideal case, the $S_p$ and the number of unit cores employed are linearly related.

\subsubsection{Strong Scaling Results and Discussions}

\begin{table}[ht!]	
\begin{center}
		\begin{tabular}{|c|c|c|c|c|c|c|}\hline
		 Cores & MUMPS & $S_p$ &SSE &FGMRES-GMRES & $S_p$ &SSE\\\hline
		 5  & 5597.97s   &      &    &273.96s  &  &\\\hline 
    	 10 & 2542.43s   & 2.20 & 110\%   & 122.14s  & 2.24  & 112\%\\ \hline
		 15 & 1822.59s   & 3.07 & 102\%   &  87.19s  & 3.14 &  105\%\\ \hline
		 20 & 1520.46s   & 3.68 & 92\%   & 69.43s  & 3.95 &  99\%   \\\hline
		 25 & 1324.26s   & 4.23 & 85\%   & 59.35s & 4.62 &  92\%\\\hline
		 30 & 1171.39s   & 4.78 & 80\%   & 49.35s   & 5.55& 93\%\\\hline
		 35 & 1101.91s   & 5.08 & 73\%   & 48.23s  & 5.68 & 81\%\\\hline
		 40 & 1197.96s   & 4.67 & 58\%   & 42.21s  & 6.49 & 81\%\\\hline
		\end{tabular}\caption{Strong scaling: T-1 model with dofs $=5415890$.}\label{t-1ss_table}
	\end{center}
\end{table}

\begin{figure}
    \centering
    \begin{minipage}{0.45\textwidth}
        \centering
        \includegraphics[width=1.0\textwidth,height=0.75\textwidth]{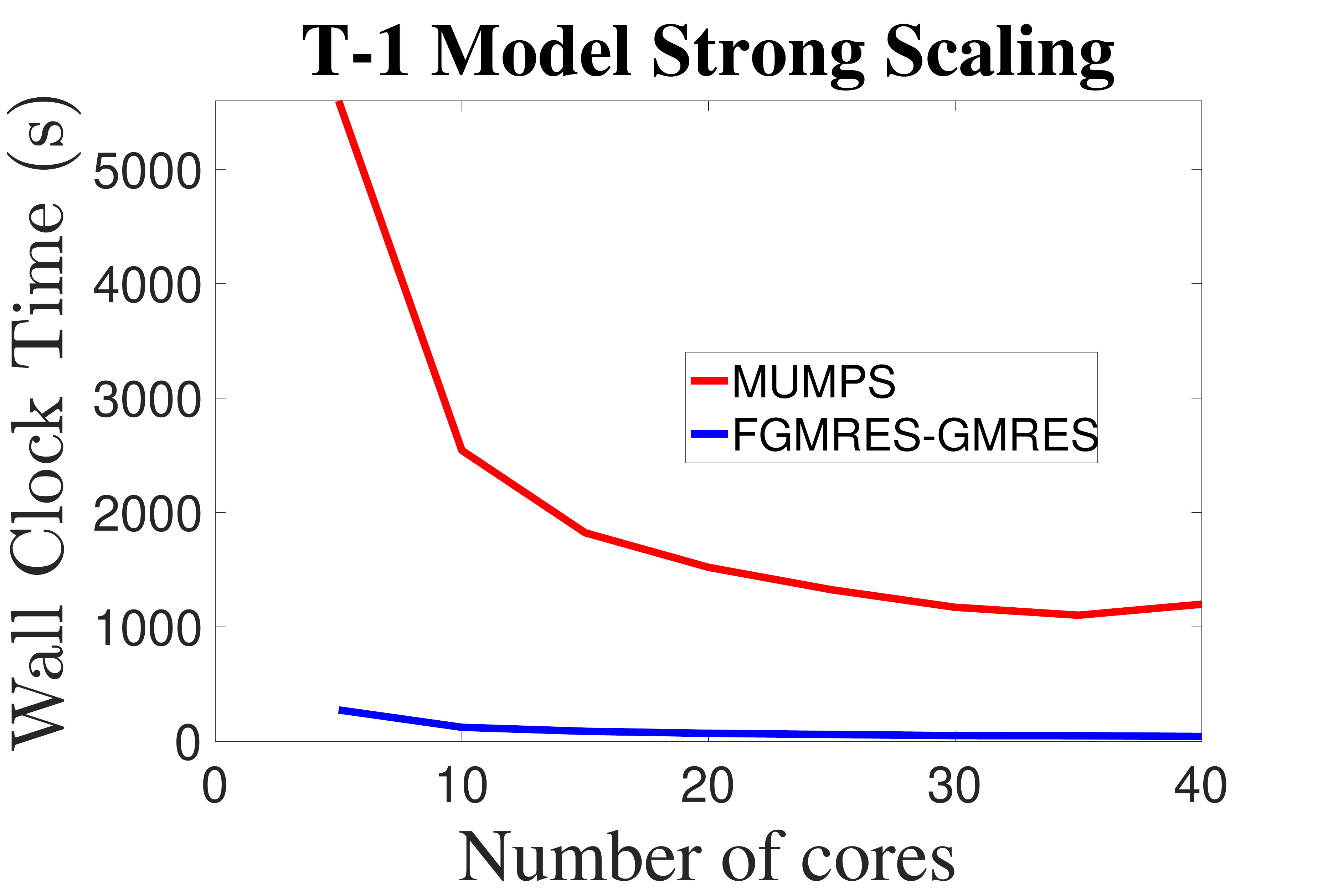} 
        \caption{Measured computational time versus number of cores.}\label{t-1ss_time}
    \end{minipage}\hfill
    \begin{minipage}{0.45\textwidth}
        \centering
        \includegraphics[width=1.0\textwidth,height=0.75\textwidth]{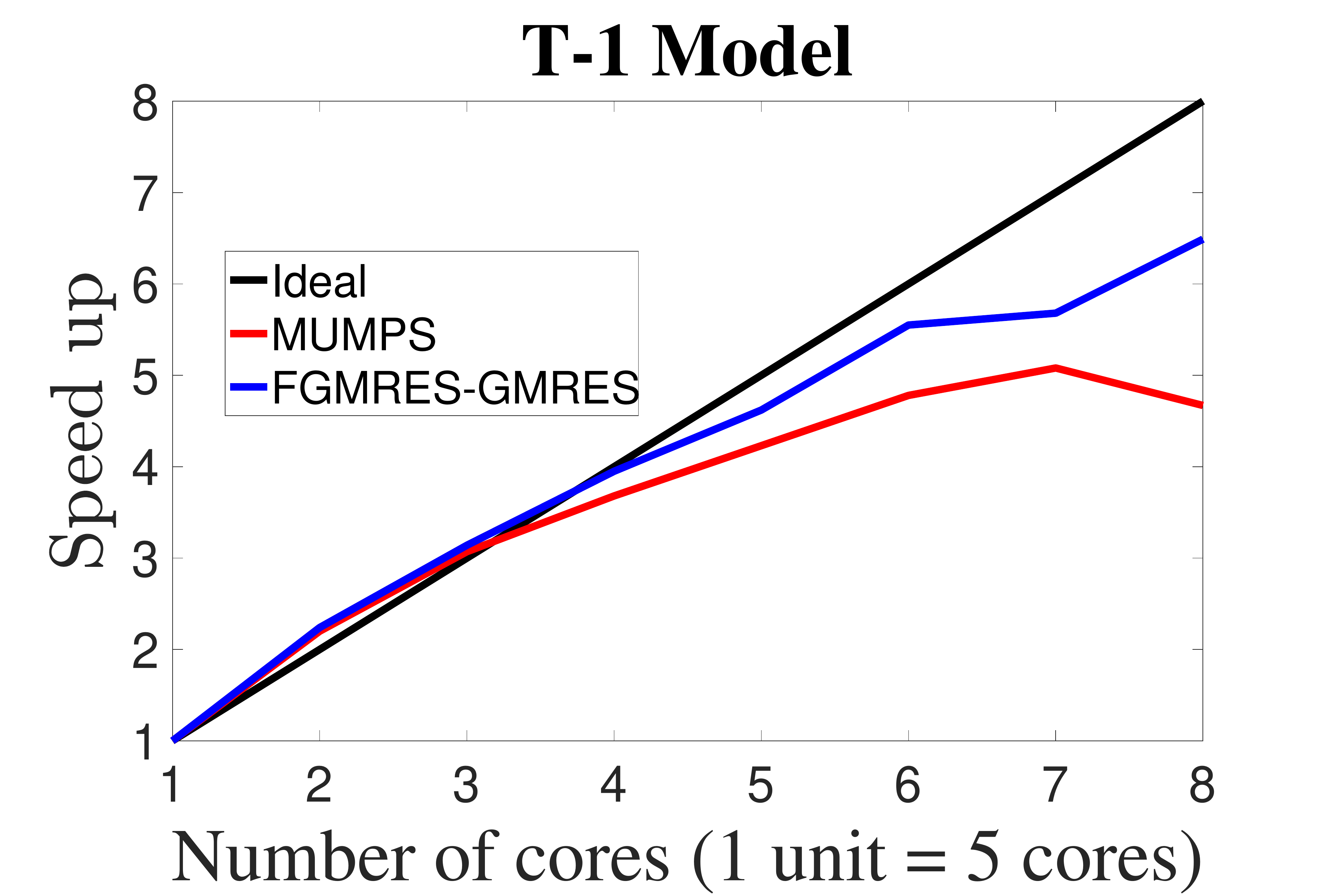} 
        \caption{Strong scaling for direct and iterative solver, speedup versus number of cores.}\label{t-1ss_speedup}
    \end{minipage}
\end{figure}

In Table \ref{t-1ss_table} we represent the results of the strong scaling with the T-1 model. In this case, we monitor the computational time corresponds to a mesh that provides 5415890 dofs, while progressively increasing the number of cores. Here, one unit core has 5 cores. To test the scalability, we vary the number of units of cores as 1, 2, 3, 4, 5, 6, 7, and 8 which are corresponding to 5, 10, 15, 20, 25, 30, 35, and 40 cores. As shown in Fig. \ref{t-1ss_speedup}, the iterative solver speedup scales almost linearly up to 30 cores, having about 180,000 dofs per core, whereas the direct solver speedup scales almost linearly up to 20 cores, having about 270,000 dofs. After the linear scalability, even the parallel performances of both solvers reduce, the iterative solver speedup remains bigger than that of the direct solver. We observe that the direct solver strong scaling efficiency drops to 58\% whereas the iterative solver drops to 81\%. In Fig. \ref{t-1ss_time}, we plotted the solving time versus the number of cores for both the solvers, which shows how faster the problem can be solved by the direct and the iterative solver. The direct solver computational time is much higher than that of the iterative solver.

\begin{figure}
    \centering
    \begin{minipage}{0.45\textwidth}
        \centering
        \includegraphics[width=1.0\textwidth,height=0.75\textwidth]{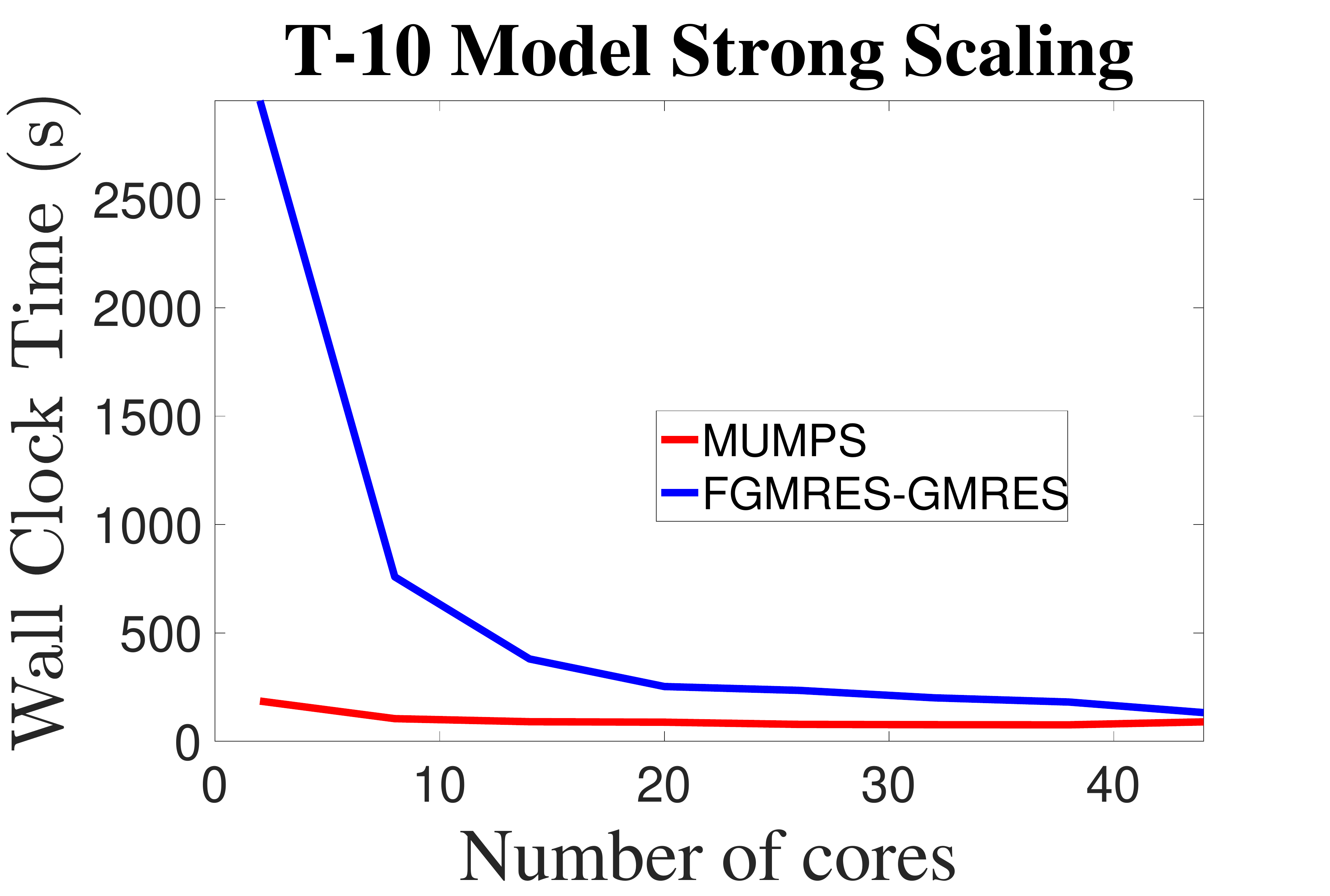} 
        \caption{Measured computational time versus number of cores.}\label{t-10ss_time}
    \end{minipage}\hfill
    \begin{minipage}{0.45\textwidth}
        \centering
        \includegraphics[width=1.0\textwidth,height=0.75\textwidth]{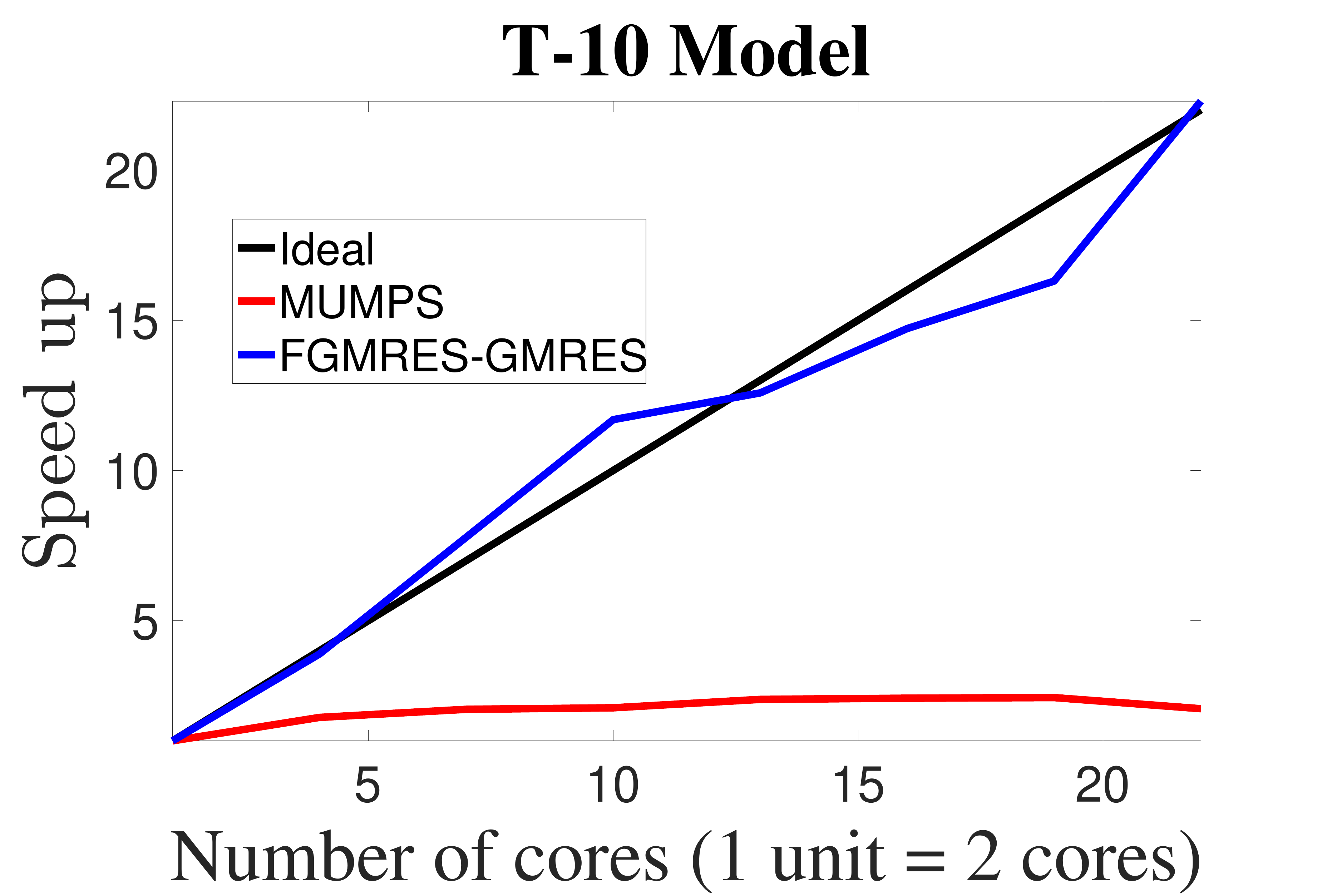} 
        \caption{Strong scaling for direct and iterative solver, speedup versus number of cores.}\label{t-10ss_speedup}
    \end{minipage}
\end{figure}

\begin{table}[ht!]	
\begin{center}
		\begin{tabular}{|c|c|c|c|c|c|c|}\hline
		 Cores & MUMPS & $S_p$ &SSE &FGMRES-GMRES & $S_p$ &SSE\\\hline
	 2   & 185.67s  &   &  &  2954.94s    &  & \\\hline
	 8& 104.58s &1.78&44.50\%&758.65s&3.89& 97.25\%\\\hline
	 14 & 90.51s &2.05&29.29\%&379.89s&7.78& 111.14\%\\\hline
	 20 & 88.21s &2.10&21.00\%&252.72s&11.69& 116.90\%\\\hline
	 26 &78.16s &2.38& 18.31\%&234.97s&12.58& 96.77\%\\\hline
	 32 & 76.61s &2.42&15.13\%&200.72s&14.72& 92.00\%\\\hline
	 38 & 76.02s &2.44&12.84\%&181.30s&16.30& 85.79\%\\\hline
	 44 &89.77s&2.07&9.41\%& 132.54s&22.29& 101.32\%
	\\\hline
	
		\end{tabular}\caption{Strong scaling: T-10 model with dofs $=1860729$.}\label{t-10ss_table}
	\end{center}
\end{table}
To study the strong scaling of the direct and iterative solver with the T-10 model, we consider a mesh that gives a total of 1860729 dofs, and solve the problem using 2, 8, 14, 20, 26, 32, 38, and 44 cores. The recorded computational time, computed speedup and SSE are presented in Table \ref{t-10ss_table}. We observe that as the number of cores increases, the iterative solver SSE does not decrease much, however, the direct solver SSE drops significantly. We plotted the computational time versus the number of cores employed in Fig. \ref{t-10ss_time}. From Fig. \ref{t-10ss_time} we observe that the direct solver shows a poor scaling while the iterative solver shows a good scaling. Since as we increase the number of cores, the direct solver fails to reduce the solving time significantly, whereas the iterative solver solving time drops exponentially. However, the iterative solver solving time is higher than that of the director solver, this is supported by the weak scaling, and since the dofs under consideration is not too high. 

We also plotted the \textit{speedup} versus the number of \textcolor{black}{processor} cores \textcolor{black}{employed} graph for both solvers and the ideal case scenario in Fig. \ref{t-10ss_speedup}. The iterative solver speedup graph changes almost linearly as the number of cores varies, which also supports the good scaling of the iterative solver, and clearly the graph shows a poor scaling phenomenon of the direct solver.

	\begin{table}[ht!]
	\begin{center}
		\begin{tabular}{|c|c|c|c|c|c|c|}\hline
		 Cores     & MUMPS & $S_p$ &SSE & FGMRES-GMRES & $S_p$ &SSE\\\hline
		 1  & 346.25s  & &&  4748.42s &&  \\\hline 
		 5  & 187.83s  & 1.84 &36.87\% &  1228.23s &  3.87& 77.32\%\\ \hline
		 10  & 140.66s  & 2.46 & 24.62\% &   783.14s & 6.06& 60.63\% \\ \hline
		 15   &   121.75s & 2.84 & 18.96\%& 551.07s & 8.62& 57.44\%    \\\hline
		20   &  133.25s  &  2.60 & 12.99\%   &  483.96s     &  9.81& 49.06\% \\\hline
		
		\end{tabular}\caption{Strong scaling: T-20 model with dofs $=1349812$.}\label{t-20ss_table_new}
	\end{center}
	\end{table}

\begin{figure}
    \centering
    \begin{minipage}{0.45\textwidth}
        \centering
        \includegraphics[width=1.0\textwidth,height=0.75\textwidth]{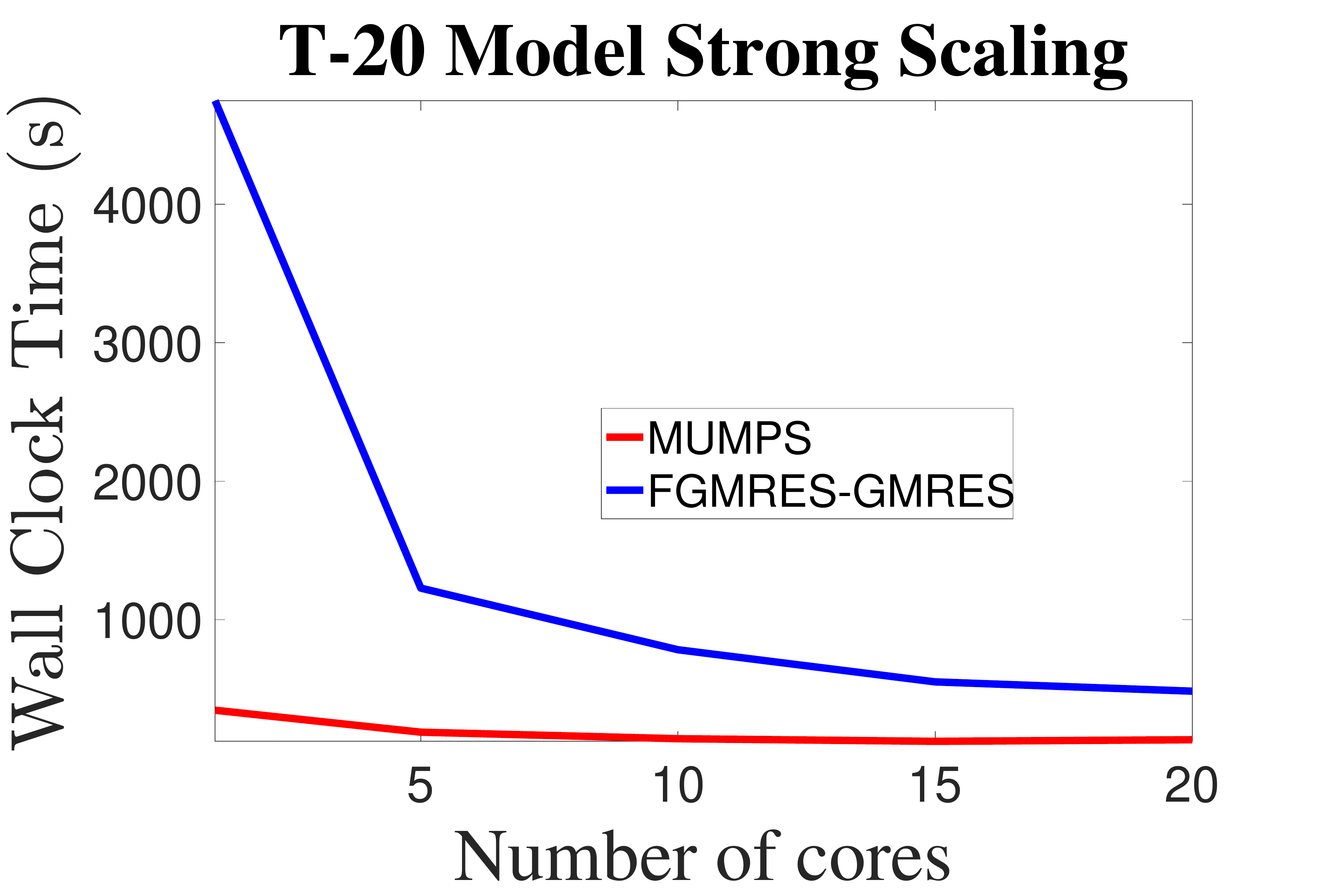} 
        \caption{Measured computational time versus number of cores.}\label{t-20ss_time}
    \end{minipage}\hfill
    \begin{minipage}{0.45\textwidth}
        \centering
        \includegraphics[width=1.0\textwidth,height=0.75\textwidth]{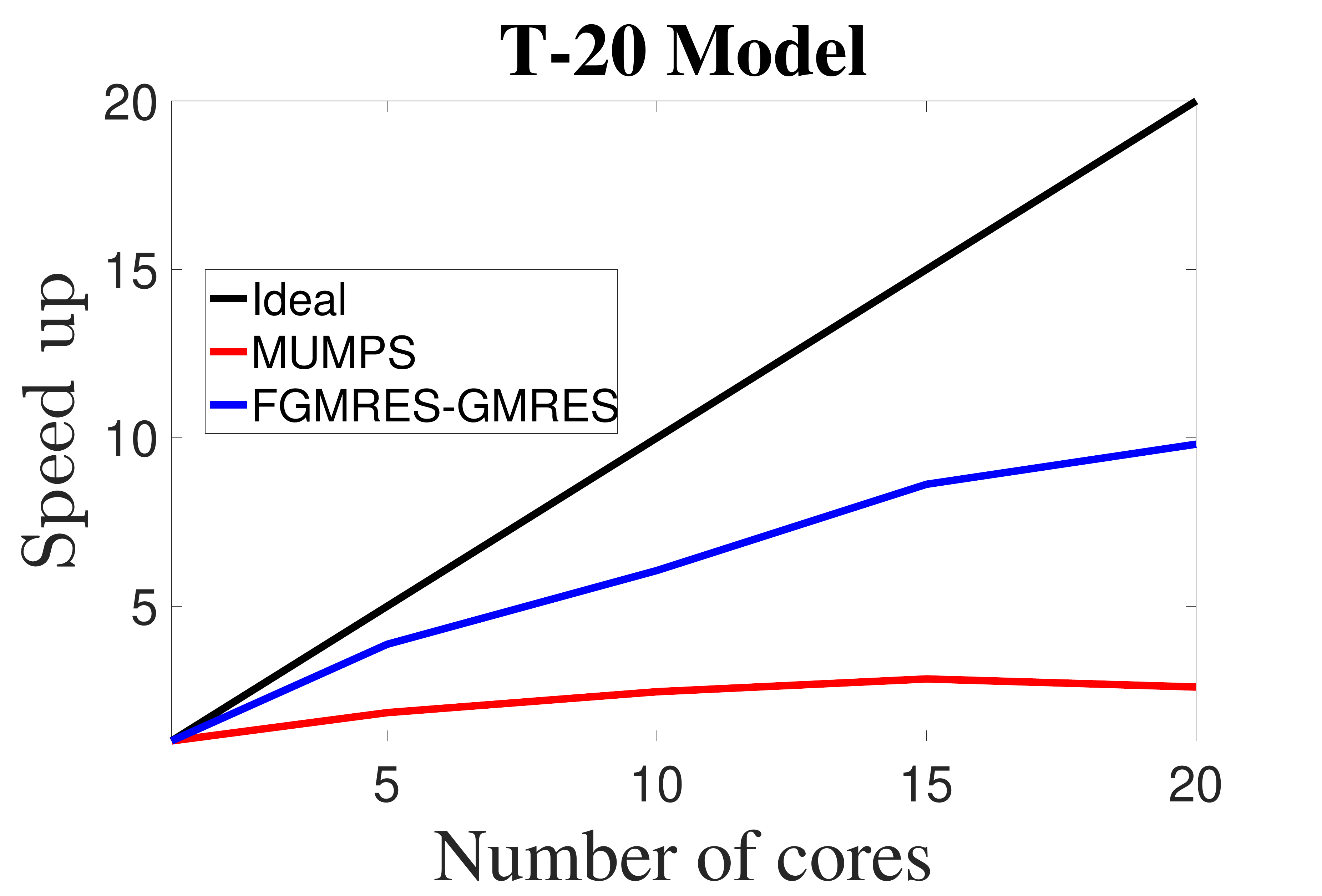} 
        \caption{Strong scaling for direct and iterative solver, speedup versus number of cores.}\label{t-20ss_speedup}
    \end{minipage}
\end{figure}

We monitor the strong scaling of the two solvers with the T-20 model and represent the data in Table \ref{t-20ss_table_new}. A mesh that provides a total of 1349812 dofs is considered. The problem is solved using 1, 5, 10, 15, and 20 cores, and while SSE drops to 12.99\% and 49.06\%, for the direct and iterative solvers, respectively. The computational time \textcolor{black}{for a single time step solve versus the number of cores employed relation is plotted in Fig. \ref{t-20ss_time}.} \textcolor{black}{We observe that as we progressively increase the number of cores, the iterative solver computational time shows an exponential drop and while the direct solver computational time drop is not that much. That is, for the T-20 model the iterative solver shows a better strong scaling than that of the direct solver.}

\textcolor{black}{We also plot the} $speedup$ versus the number of cores employed relation in Fig. \ref{t-20ss_speedup}. Clearly, the iterative solver speedup remains bigger than that of the direct solver \textcolor{black}{for all our experiments. For the direct solver the maximum $S_p$ we found is 2.84 using 15 cores if we further increase the number of cores to 20, the $S_p$ decreases to 2.60. That is, in all the scenarios we considered, the direct solver solved the problem 2.84 times faster than the direct solver with a single core. On the other hand, the $S_p$ for the iterative solver increases monotonically as we progressively increase the number of cores, and the maximum $S_p$ we found is 9.81 which corresponds to 20 cores. That is, the iterative solver solves the problem 9.81 times faster using 20 cores than it does using a single core.} 

\begin{table}[ht!]	
\begin{center}
		\begin{tabular}{|c|c|c|c|c|c|c|}\hline
		 Cores & MUMPS & $S_p$ &SSE &FGMRES-GMRES & $S_p$ &SSE\\\hline
		 8 & 1420.11s & &&7455.98s&& \\\hline
		 16 & 1097.91s&1.29&64.67\%&3658.53s& 2.04&101.90\%\\\hline
		 32 & 885.91s& 1.60& 40.07\%& 2228.31s &3.35&83.65\%\\\hline
		 96 & 809.37s&1.75&14.62\%& 1147.35s&6.50& 54.15\%\\\hline
    	 192   & 770.10s  &1.84   &7.68\%  & 751.43s & 9.92&41.34\%\\\hline
    	 256 & 788.62s&1.80 &5.63\% & 720.42s&10.35 &32.34\% \\\hline
    	 288 & 744.80s &1.91&5.30\% &683.46s&10.91&30.30\%\\\hline
    	 304& 780.06s&1.82&4.79\%&668.46s&11.15&29.35\%\\\hline
		\end{tabular}\caption{Strong scaling: T-30 model with dofs $=10479901$.}\label{t-30ss_table}
	\end{center}
\end{table}

\begin{figure}
    \centering
    \begin{minipage}{0.45\textwidth}
        \centering
        \includegraphics[width=1.0\textwidth,height=0.75\textwidth]{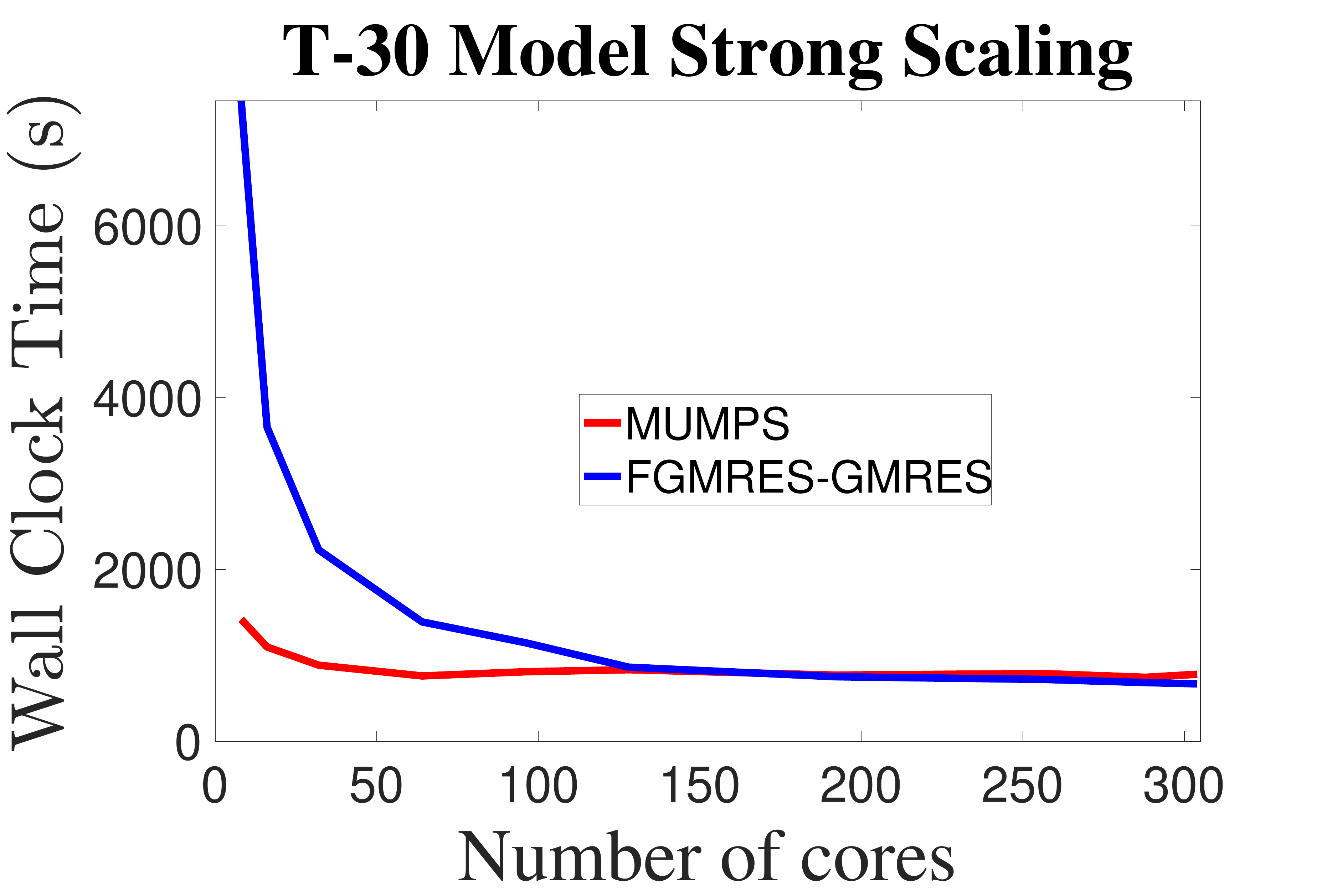} 
        \caption{Measured computational time versus number of cores.}\label{t-30ss_time}
    \end{minipage}\hfill
    \begin{minipage}{0.45\textwidth}
        \centering
        \includegraphics[width=1.0\textwidth,height=0.75\textwidth]{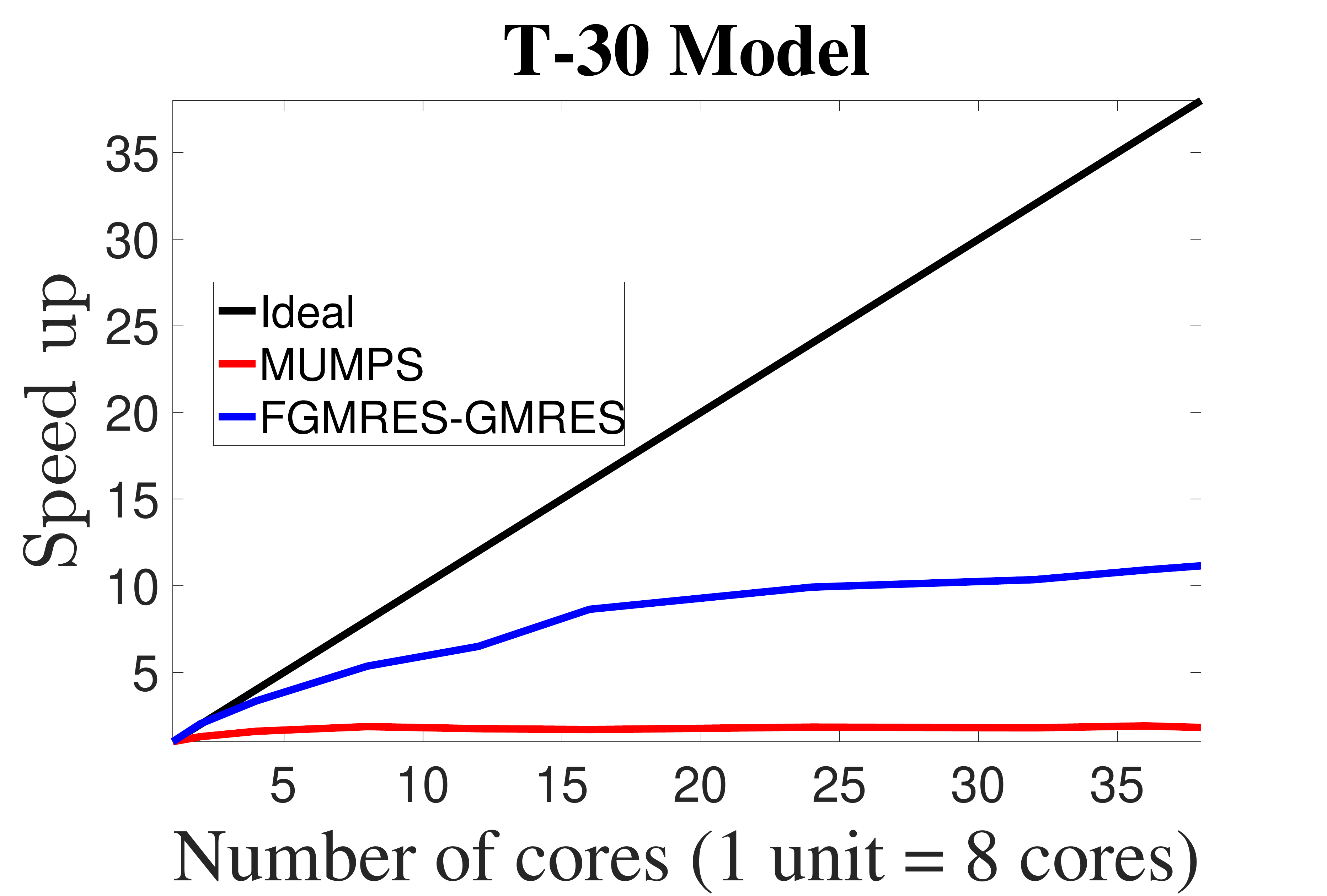} 
        \caption{Strong scaling for direct and iterative solver, speedup versus number of cores.}\label{t-30ss_speedup}
    \end{minipage}
\end{figure}

Finally, we investigate the strong scaling of the direct and iterative solvers with the T-30 model. In this case, we consider a mesh which provides a total of 10479901 dofs. We solve the problem employing 8, 16, 32, 96, 192, 256, 288, and 304 cores using both the direct and iterative solver for a single time step and records the computational time in Table \ref{t-30ss_table}. In this case, $C_1(10479901)=8$, and $p$ varies as 2, 4, 12, 24, 32, 36, and 38. We compute $S_p$ and SSE for both the solvers. From our observations, we found the maximum $S_p$ equals to 1.91 and 11.15 for the direct and iterative solver, respectively. That is, we solve the problem 1.91 times faster by the direct solver and 11.15 times faster by the iterative solver progressively increasing the number of cores. We plot the $speedup$ versus the number of employed cores for the T-30 model in Fig. \ref{t-30ss_speedup}. We observe the iterative solver $speedup$ is higher than the direct solver $speedup$.
\textcolor{black}{
We plot the computing time versus the number of cores employed for both solvers in Fig. \ref{t-30ss_time}. We observe a drastic drop in the computational time with the iterative solver than that with the direct solver. This indicates the iterative solver provides a better scaling than the direct solver.}
\section{Conclusion}\label{conclusion-futuredir}
Numerical solutions of complex PDEs are often one of the most challenging tasks, which involve finding solutions of systems of linear equations with multi-billions degrees of freedom. \textcolor{black}{In this paper, we have presented a voltage excited magnetic field formulation for the charging simulation of non-insulated superconducting pancake solenoids.} \textcolor{black}{The numerical difficulty for the simulation of the problem arises from the fact that the HTS coil is very thin but with excessively high conductivity whereas its surrounding materials have the opposite properties.}

\textcolor{black}{In this paper, we have explored an iterative solver for Maxwell's equation simulation, which is a combination FGMRES and GMRES solver with a parallel AMS Solver preconditioner.}
We have investigated the scalability performance of the direct solver, MUMPS, and the iterative solver, FGMRES-GMRES with AMS preconditioner, on 1, 10, 20, and 30 turns pancake solenoids. \textcolor{black}{The direct solver is efficacious for problems with fewer degrees of freedom and provides an additional advantage when the conductivity of the diffusion equation is independent of time. Since in this case, \textcolor{black}{the system matrix remains same for all time step but the right hand side vector changes and the direct solver} factorizes the system matrix only once and repeatedly use them for the further timesteps.}
\textcolor{black}{On the other hand,} the iterative solver is faster, uses less memory, and scale better than the direct solver, especially to the problem that has complex geometry and/or higher degrees of freedom. \textcolor{black}{This iterative solver along with the AMS preconditioner can successfully solve the problem of size with at least 48 million degrees of freedom.}

\textcolor{black}{As a benchmark, we have computed the Biot-Savart fields and compared them with the FEM results, and found excellent agreement.}
\textcolor{black}{It is observed that if we increase the number of turns in a charging solenoid, the charging time of the model and the computational time of the iterative solver both increase. Finally, we have also found that as the conductivity of the superconductor increases the problem becomes harder and the iterative solver needs longer time to converge.}

\textcolor{black}{We believe the most important future research avenue would be the incorporation of quench analysis \cite{bhattarai2020understanding} in our mathematical model and explore the charging behavior of non-insulated superconducting pancake solenoids. We will use the proposed efficient solver, FGMRES-GMRES, for problems with large degrees of freedom, and will continue to investigate further for a better preconditioner.}  

\textcolor{black}{Apart from the voltage excitation problem, we plan to investigate a current intensity excitation modeling \cite{rodriguez2010eddy} and explore the charging behavior ramping up the current slowly. In a current intensity excitation problem, the voltage remains an unknown.}

\textcolor{black}{Instead of $\bH$-formulation, we plan to investigate the eddy current simulation on a multiply connected region based on $\bH-\Phi$ field formulation following the work in \cite{smajic2019treatment}. The magnetic field $\bH$ will be computed using the iterative solver in the conducting region and the multi-valued magnetic scalar potential $\Phi$ will be the solution of a diffusion BVP in the non-conduction region. Finally, both solutions will be assembled together using the appropriate tangential continuity condition across the interfaces between different tetrahedral elements. This approach will potentially reduce the cost of computing the magnetic field in the air domain.}

\textcolor{black}{We plan to explore the stability and error estimate analysis for the fully discrete scheme used in this paper following the works in \cite{akbas2016numerical, heister2017decoupled}.}
\textcolor{black}{The study of ensemble magnetic field calculations \cite{mohebujjaman2017efficient} with the advantage of block linear solver can also be a new research avenue. We also want to explore the reduced order modeling of the magnetic field simulation in a non-insulated superconducting pancake solenoids following the analysis and experiments given in \cite{mohebujjaman2017energy}.}

\vspace{2mm}
\textbf{Acknowledgment.} The work was supported by Commonwealth Fusion Systems grant RPP002. \textcolor{black}{We thank Dan Brunner for his constructive comments and suggestions, which significantly improved the quality of the manuscript.}
\vspace{2mm}
\appendix


\section{Biot-Savart Field Computation}
For the spiral models (T-10, T-20, and T-30), we assume $x$ be the uniform gap between two turns, and the innermost radius of the filament is $r$. Then the radius of a spiral changes as
\begin{align}
    R=r+\frac{x}{2\pi}\theta,\hspace{2mm}0\leq\theta\leq 2n\pi,
\end{align}
where $n\in\mathbb{N}$, represents the number of turns. The position vector that represents the spiral filament, can be written as
\begin{align}
\bs(\theta)=<R\cos{\theta},R\sin{\theta},d>.  
\end{align}

Therefore, the magnetic field at the origin produced by the current flow in the lower pancake spiral filament is given by
\begin{align}
    \bB_L(\vec{0})=-\frac{\mu_0I}{4\pi}\begin{pmatrix}\bigint_0^{2n\pi}\frac{\frac{xd}{2\pi}\sin\theta+d\left(r+\frac{x}{2\pi}\theta\right)\cos{\theta}}{\big\{\left(r+\frac{x}{2\pi}\theta\right)^2+d^2\big\}^{3/2}}d\theta\\\\
    \bigint_0^{2n\pi}\frac{-\frac{xd}{2\pi}\cos{\theta}+d\left(r+\frac{x}{2\pi}\theta\right)\sin{\theta}}{\big\{\left(r+\frac{x}{2\pi}\theta\right)^2+d^2\big\}^{3/2}}d\theta\\\\
   \bigint_0^{2n\pi} \frac{-(r+\frac{x}{2\pi}\theta)^2}{\big\{\left(r+\frac{x}{2\pi}\theta\right)^2+d^2\big\}^{3/2}}d\theta
    \end{pmatrix}.\label{BS-analytic}
\end{align}

Since the upper pancake lies on the $xy$-plane, we have in this case $d=0$, and thus the magnetic field contribution at the origin, is given by
\begin{align}
    \bB_U(\vec{0})=<0,0,\frac{\mu_0I}{2x}\ln\left(1+\frac{nx}{r}\right)>.
\end{align}

Finally, we consider the vector that represents the connection line between the two pancakes, is given by
\begin{align*}
    \bs_{con}=<r,0, z>, \hspace{2mm}\text{where}\hspace{2mm}d\le z\le 0,
\end{align*}
and compute the magnetic field at the origin as
\begin{align}
    \bB_{con}(\vec{0})=\frac{\mu_0I}{4\pi}<0,\frac{-d}{\sqrt{r^2+d^2}},0>.
\end{align}

The resultant magnetic field at the origin is then given by
\begin{align}
    \bB(\vec{0})=\bB_L(\vec{0})+\bB_U(\vec{0})+\bB_{con}(\vec{0}),
\end{align}
\textcolor{black}{and thus its strength is $\|\bB(\vec{0})\|$.}
 The fully charged magnetic field strength \textcolor{black}{at the origin }  $B_0^{max}=\|\bB(\vec{0})\|$, \textcolor{black}{when $\bB(\vec{0})$ reaches its statistically steady-state}. The computation of $B_0^{max}$ for the T-1 model using the Biot-Savart law is straight forward, and thus omitted. 


\bibliographystyle{plain}
\bibliography{main}

\begin{thebibliography}{10}

\bibitem{mfem-library}
{MFEM}: Modular finite element methods library.
\newblock \url{https://mfem.org/}.

\bibitem{akbas2016numerical}
M.~Akbas, S.~Kaya, M.~Mohebujjaman, and L.~Rebholz.
\newblock Numerical analysis and testing of a fully discrete, decoupled
  penalty-projection algorithm for mhd in els{\"a}sser variable.
\newblock {\em Int. J. Numer. Anal. Model}, 13(1):90--113, 2016.

\bibitem{amestoy2000multifrontal}
P.~R. Amestoy, I.~S. Duff, and J.~L$\ensuremath{'}$Excellent.
\newblock Multifrontal parallel distributed symmetric and unsymmetric solvers.
\newblock {\em Computer methods in applied mechanics and engineering},
  184(2-4):501--520, 2000.

\bibitem{amestoy2000mumps}
P.~R. Amestoy, I.~S. Duff, J.~L$\ensuremath{'}$Excellent, and J.~Koster.
\newblock {MUMPS}: {A} general purpose distributed memory sparse solver.
\newblock {\em Sørevik T., Manne F., Gebremedhin A.H., Moe R. (eds) Applied
  Parallel Computing. New Paradigms for HPC in Industry and Academia. PARA
  2000. Lecture Notes in Computer Science}, 1947:121--130, 2000.

\bibitem{bangerthLecture34}
W.~Bangerth.
\newblock Finite element methods in scientific computing, {L}ecture 34: What
  solver to use.
\newblock
  \url{https://www.math.colostate.edu/~bangerth/videos/676/slides.34.pdf}.

\bibitem{bhattarai2020understanding}
K.~R. Bhattarai, K.~Kim, K.~Kim, K.~Radcliff, X.~Hu, C.~Im, T.~Painter,
  I.~Dixon, D.~Larbalestier, S.~Lee, and S.~Hahn.
\newblock Understanding quench in no-insulation (ni) rebco magnets through
  experiments and simulations.
\newblock {\em Superconductor Science and Technology}, 33(3):035002, 2020.

\bibitem{biro1999edge}
O.~B{\'\i}r{\'o}.
\newblock Edge element formulations of eddy current problems.
\newblock {\em Computer methods in applied mechanics and engineering},
  169(3-4):391--405, 1999.

\bibitem{bossavit2000most}
A.~Bossavit.
\newblock Most general `non-local' boundary conditions for the maxwell equation
  in a bounded region.
\newblock {\em COMPEL}, 19(2):239--245, 2000.

\bibitem{Brandt1996Superconductors}
E.~H. Brandt.
\newblock Superconductors of finite thickness in a perpendicular magnetic
  field: Strips and slabs.
\newblock {\em Physical Review B}, 54(6):4246--4264, 1996.

\bibitem{by201416}
S.~By, J.~V. Rispoli, S.~Cheshkov, I.~Dimitrov, J.~Cui, S.~Seiler, S.~Goudreau,
  C.~Malloy, S.~M. Wright, and M.~P. McDougall.
\newblock A 16-channel receive, forced current excitation dual-transmit coil
  for breast imaging at 7{T}.
\newblock {\em PloS one}, 9(11), 2014.

\bibitem{davis2004algorithm}
T.~A. Davis.
\newblock Algorithm 832: {UMFPACK} v4.3---an unsymmetric-pattern multifrontal
  method.
\newblock {\em ACM Transactions on Mathematical Software (TOMS)},
  30(2):196--199, 2004.

\bibitem{falgout2002hypre}
R.~D. Falgout and U.~M. Yang.
\newblock hypre: A library of high performance preconditioners.
\newblock {\em International Conference on Computational Science}, pages
  632--641, 2002.

\bibitem{ghysels2016efficient}
P.~Ghysels, X.~S. Li, F.~Rouet, S.~Williams, and A.~Napov.
\newblock An efficient multicore implementation of a novel hss-structured
  multifrontal solver using randomized sampling.
\newblock {\em SIAM Journal on Scientific Computing}, 38(5):S358--S384, 2016.

\bibitem{ginsberg1992physical}
D.~M. Ginsberg.
\newblock {\em Physical properties of high temperature superconductors III},
  volume~3.
\newblock World Scientific, 1992.

\bibitem{hahn2010hts}
S.~Hahn, D.~K. Park, J.~Bascunan, and Y.~Iwasa.
\newblock {HTS} pancake coils without turn-to-turn insulation.
\newblock {\em IEEE transactions on applied superconductivity},
  21(3):1592--1595, 2010.

\bibitem{he2020efficient}
B.~He, C.~Lu, N.~Chen, D.~Lin, and P.~Zhou.
\newblock An efficient parallel computing method for the steady-state analysis
  of electric machines using the woodbury formula.
\newblock {\em IEEE Transactions on Magnetics}, 56(2):1--4, 2020.

\bibitem{heister2017decoupled}
T.~Heister, M.~Mohebujjaman, and L.~G. Rebholz.
\newblock Decoupled, unconditionally stable, higher order discretizations for
  mhd flow simulation.
\newblock {\em Journal of Scientific Computing}, 71(1):21--43, 2017.

\bibitem{hestenes1952methods}
M.~R. Hestenes and E.~Stiefel.
\newblock Methods of conjugate gradients for solving linear systems.
\newblock {\em Journal of research of the National Bureau of Standards},
  49(6):409--436, 1952.

\bibitem{hiptmair2006auxiliary}
R.~Hiptmair, G.~Widmer, and J.~Zou.
\newblock Auxiliary space preconditioning in $\textit{{\textbf{H}}}_0(curl;
  \mathrm{\Omega})$.
\newblock {\em Numerische Mathematik}, 103(3):435--459, 2006.

\bibitem{hiptmair2008auxiliary}
R.~Hiptmair and J.~Xu.
\newblock Auxiliary space preconditioning for edge elements.
\newblock {\em IEEE Transactions on Magnetics}, 44(6):938--941, 2008.

\bibitem{huo2020designing}
Z.~Huo, G.~Mei, G.~Casolla, and F.~Giampaolo.
\newblock Designing an efficient parallel spectral clustering algorithm on
  multi-core processors in julia.
\newblock {\em Journal of Parallel and Distributed Computing}, 2020.

\bibitem{kolev2009parallel}
T.~V. Kolev and P.~S. Vassilevski.
\newblock Parallel auxiliary space {AMG} for \textbf{{H}} (curl) problems.
\newblock {\em Journal of Computational Mathematics}, pages 604--623, 2009.

\bibitem{kuzmin2013fast}
A.~Kuzmin, M.~Luisier, and O.~Schenk.
\newblock Fast methods for computing selected elements of the green’s
  function in massively parallel nanoelectronic device simulations.
\newblock In {\em European Conference on Parallel Processing}, pages 533--544.
  Springer, 2013.

\bibitem{langer2018direct}
U.~Langer and M.~Neum{\"u}ller.
\newblock Direct and iterative solvers.
\newblock In {\em Computational Acoustics}, pages 205--251. Springer, 2018.

\bibitem{li2011superlu}
X.~S. Li, J.~Demmel, J.~Gilbert, L.~Grigori, and M.~Shao.
\newblock Super{LU}.
\newblock {\em Encyclopedia of Parallel Computing}, pages 1955--1962, 2011.

\bibitem{liu2017analysis}
D.~Liu, H.~Yong, and Y.~Zhou.
\newblock Analysis of charging and sudden-discharging characteristics of
  no-insulation rebco coil using an electromagnetic coupling model.
\newblock {\em AIP Advances}, 7(11):115104, 2017.

\bibitem{meng2020effect}
S.~Meng, B.~Zhu, Z.~Zhuang, X.~Chen, and C.~Tang.
\newblock Effect of excitation coil voltage on {T}i{A}l{S}i{N} coating on
  42crmo steel surface.
\newblock {\em Materials Research Express}, 7(5):056519, 2020.

\bibitem{mohebujjaman2017efficient}
M.~Mohebujjaman and L.~G. Rebholz.
\newblock An efficient algorithm for computation of mhd flow ensembles.
\newblock {\em Computational Methods in Applied Mathematics}, 17(1):121--137,
  2017.

\bibitem{mohebujjaman2017energy}
M.~Mohebujjaman, L.~G. Rebholz, X.~Xie, and T.~Iliescu.
\newblock Energy balance and mass conservation in reduced order models of fluid
  flows.
\newblock {\em Journal of Computational Physics}, 346:262--277, 2017.

\bibitem{ono2020scalable}
K.~Ono, T.~Kato, S.~Ohshima, and T.~Nanri.
\newblock Scalable direct-iterative hybrid solver for sparse matrices on
  multi-core and vector architectures.
\newblock In {\em Proceedings of the International Conference on High
  Performance Computing in Asia-Pacific Region}, pages 11--21, 2020.

\bibitem{paige1975solution}
C.~C. Paige and M.~A. Saunders.
\newblock Solution of sparse indefinite systems of linear equations.
\newblock {\em SIAM journal on numerical analysis}, 12(4):617--629, 1975.

\bibitem{Prigozhin1996Bean}
L.~Prigozhin.
\newblock The bean model in superconductivity: Variational formulation and
  numerical solution.
\newblock {\em Journal of Computational Physics}, 129(1):190--200, 1996.

\bibitem{rodriguez2010eddy}
A.~A. Rodr{\'\i}guez and A.~Valli.
\newblock {\em Eddy current approximation of Maxwell equations: {T}heory,
  algorithms and applications}, volume~4.
\newblock Springer Science \& Business Media, 2010.

\bibitem{rouet2016distributed}
F.~Rouet, X.~S. Li, P.~Ghysels, and A.~Napov.
\newblock A distributed-memory package for dense hierarchically semi-separable
  matrix computations using randomization.
\newblock {\em ACM Transactions on Mathematical Software (TOMS)}, 42(4):1--35,
  2016.

\bibitem{ruiz2004computer}
D.~Ruiz-Alonso, T.~Coombs, and A.~M. Campbell.
\newblock Computer modelling of high-temperature superconductors using an
  {A}--{V} formulation.
\newblock {\em Superconductor Science and Technology}, 17(5):S305, 2004.

\bibitem{saad1993flexible}
Y.~Saad.
\newblock A flexible inner-outer preconditioned gmres algorithm.
\newblock {\em SIAM Journal on Scientific Computing}, 14(2):461--469, 1993.

\bibitem{saad1986gmres}
Y.~Saad and M.~H. Schultz.
\newblock {GMRES}: A generalized minimal residual algorithm for solving
  nonsymmetric linear systems.
\newblock {\em SIAM Journal on scientific and statistical computing},
  7(3):856--869, 1986.

\bibitem{Shiraiwa2017RF}
S.~Shiraiwa, J.~C. Wright, P.~T. Bonoli, T.~Kolev, and M.~Stowell.
\newblock R{F} wave simulation for cold edge plasmas using the {MFEM} library.
\newblock {\em EPJ Web of Conferences}, 157(03048), 2017.

\bibitem{shoukourian2014predicting}
H.~Shoukourian, T.~Wilde, A.~Auweter, and A.~Bode.
\newblock Predicting the energy and power consumption of strong and weak
  scaling hpc applications.
\newblock {\em Supercomputing frontiers and innovations}, 1(2):20--41, 2014.

\bibitem{simoncini2003flexible}
V.~Simoncini and D.~B. Szyld.
\newblock Flexible inner-outer krylov subspace methods.
\newblock {\em SIAM Journal on Numerical Analysis}, 40(6):2219--2239, 2003.

\bibitem{smajic2019treatment}
J.~Smajic, M.~K. Bucher, C.~J{\"a}ger, and R.~Christen.
\newblock Treatment of multiply connected domains in time-domain discontinuous
  galerkin ${\bH}-{\Phi}$ eddy current analysis.
\newblock {\em IEEE Transactions on Magnetics}, 55(6):1--4, 2019.

\bibitem{sun1994scalability}
X.~Sun and D.~T. Rover.
\newblock Scalability of parallel algorithm-machine combinations.
\newblock {\em IEEE Transactions on parallel and Distributed Systems},
  5(6):599--613, 1994.

\bibitem{van1992bi}
H.~A. Van~der Vorst.
\newblock Bi-{CGSTAB}: A fast and smoothly converging variant of {Bi-CG} for
  the solution of nonsymmetric linear systems.
\newblock {\em SIAM Journal on scientific and Statistical Computing},
  13(2):631--644, 1992.

\end{thebibliography}
\end{document}